\documentclass{aa}
\def\puncspace{\ifmmode\,\else{\ifcat.\C{\if.\C\else\if,\C\else\if?\C\else%
\if:\C\else\if;\C\else\if-\C\else\if)\C\else\if/\C\else\if]\C\else\if'\C%
\else\space\fi\fi\fi\fi\fi\fi\fi\fi\fi\fi}%
\else\if\empty\C\else\if\space\C\else\space\fi\fi\fi}\fi}
\def\SP{\let\\=\empty\futurelet\C\puncspace}

\def\h1{$h^{-1}$\SP}

\def\etal{{\it et al.\/}\ }

\def\eg{{\it e.g.\/}\rm,\ }
\def\lsim{~\rlap{$<$}{\lower 1.0ex\hbox{$\sim$}}}
\def\gsim{~\rlap{$>$}{\lower 1.0ex\hbox{$\sim$}}}
\def\void#1{{}}

\usepackage{times}
\usepackage{mathptm}
\fontfamily{ptmcm}\selectfont
\usepackage{graphicx}

\begin{document}

   \thesaurus{06     % A&A Section 6: Form. struct. and evolut. of stars
              (03.11.1;  % Cosmogony,
               16.06.1;  % Planets and satellites: general,
               19.06.1;  % Solar system: general,
               19.37.1;  % Stars: formation of,
               19.53.1;  % Stars: oscillations of,
               19.63.1)} % Stars: structure of.
   \title{ESO Imaging Survey}

   \subtitle{IV. Exploring the EIS  Multicolor Data}

\author {S. Zaggia\inst{1,2} \and I. Hook\inst{1} \and R. Mendez\inst{1,3}
\and L. da Costa\inst{1} \and L.F. Olsen\inst{1,4}
\and M. Nonino\inst{1,5}  \and A. Wicenec\inst{1} 
\and C. Benoist\inst{1,6} 
\and E. Deul\inst{1,7} \and T. Erben\inst{1,8}  
\and M.D. Guarnieri\inst{1,9} \and R. Hook\inst{10} \and I. Prandoni\inst{1,11}
\and M. Scodeggio\inst{1} \and R. Slijkhuis\inst{1,7} 
\and  R. Wichmann\inst{1,12}
}

\institute{
European Southern Observatory, Karl-Schwarzschild-Str. 2, D--85748
Garching b. M\"unchen, Germany
\and Osservatorio Astronomico di Capodimonte, via Moiariello 15,
I-80131.  Napoli, Italy
\and Cerro Tololo Inter-American Observatory, Casilla 603, La Serena, Chile
 \and
Astronomisk Observatorium, Juliane Maries Vej 30, DK-2100 Copenhagen, 
Denmark
\and Osservatorio Astronomico di Trieste, Via G.B. Tiepolo 11, I-31144
Trieste, Italy 
\and DAEC, Observatoire de
Paris-Meudon, 5 Pl. J. Janssen, 92195 Meudon Cedex, France
\and Leiden Observatory, P.O. Box 9513, 2300 RA Leiden, The
Netherlands
\and Max-Planck Institut f\"ur Astrophysik, Postfach 1523 D-85748,
 Garching b.  M\"unchen, Germany 
\and Osservatorio Astronomico di Pino Torinese, Strada Osservatorio 20, I-10025 
Torino, Italy
\and Space Telescope -- European Coordinating Facility, Karl-Schwarzschild-Str. 2, 
D--85748 Garching b. M\"unchen, Germany 
\and Istituto di Radioastronomia del
CNR, Via Gobetti 101, 40129 Bologna, Italy
\and IUCAA, Post Bag 4,
Ganeshkhind, Pune 411007, India
}

   \offprints{M. Nonino}

   \date{Received ; accepted }

   \maketitle
    
%   \today

\begin{abstract} This paper presents preliminary lists of
potentially interesting point-like sources extracted from multicolor
data obtained for a 1.7 square degree region near the South Galactic
Pole.  The region has been covered by the ESO Imaging Survey (EIS) in
$B, V$ and $I$ and offers a unique combination of area and depth.
These lists, containing a total of 330 objects nearly all brighter
than $I \sim 21.5$, over 1.27 square degrees (after removing some bad
regions), are by-products of the process of verification and quality
control of the object catalogs being produced. Among the color
selected targets are candidate very low mass stars/brown dwarfs (54),
white-dwarfs (32), and quasars (244). In addition, a probable fast
moving asteroid was identified. The objects presented here are natural
candidates for follow-up spectroscopic observations and illustrate the
usefulness of the EIS data for a broad range of science and for
providing possible samples for the first year of the VLT.

\keywords{imaging survey -- color catalog -- quasars}
\end{abstract}

%
%________________________________________________________________

\section{Introduction}

The present paper is the fourth of a series presenting the data
accumulated by the public ESO Imaging Survey (EIS), being carried out
in preparation for the first year of regular operation of the VLT. As
described in previous papers (Renzini \& da Costa 1997, Nonino \etal
1998, hereafter paper~I) the main goals of the EIS project are to
conduct an imaging survey suitable for finding ``rare'' objects for
follow-up observations with the VLT and to lay down the groundwork for
more ambitious wide-field digital, multicolor imaging surveys. The
adopted strategy was designed to search for distant clusters of
galaxies, quasars, high-redshift galaxies and to identify rare stellar
types.

As part of EIS-wide, observations were obtained in three passbands
($B, V$ and $I$) over an area of about 1.7 square degrees in a region
near the South Galactic Pole (EIS-wide patch~B). The region was
selected because of the high density of known intermediate red-shift
quasars. The multicolor data for patch B are now being released and in
a separate paper the observations, calibration and the overall quality
of the data are described (Prandoni \etal 1998, paper~III). In that
paper, preliminary single passband catalogs were presented and
evaluated by comparing the star- and galaxy-counts with models and
results from other surveys and the EIS patch A, presented in
paper~I. The good agreement found from these comparisons indicates
that the available data, reduction procedures and object catalogs
extracted in each of the passbands are reliable. Furthermore,
comparison of the observed color distribution for point-like sources,
derived from a preliminary version of a color catalog, with
predictions of galactic models also shows reasonable agreement.

Although encouraging, the above results do not fully characterize the
color catalog, in particular if it is to be used to identify different
types of objects based on color selection criteria.  Target selection
based on the location of objects in color-color space, require a
careful investigation of the performance of the pipeline since any
problem may lead to objects with spurious colors.  Colors are
sensitive to a number of effects such as the observing conditions, the
photometric and astrometric solutions in the different passbands and
the de-blending algorithm. In order to have a better understanding of
the impact of these effects in the color catalog, in this paper the
$(B-V)$ versus $(V-I)$ diagram for point-sources is used to select
objects lying in potentially interesting regions of color-color
space. These objects, examined individually, are used to generate
preliminary lists of candidates for different type of objects. The
criteria adopted are conservative and not optimal but the results
serve to illustrate the type of objects one may expect to find and the
approximate size of candidate samples for follow-up spectroscopic
observations. Since the data are public, improved selection can be
carried out by other interested groups.

The goals of the present paper are: 1) to test the performance of the
EIS pipeline in the production of object catalogs; 2) to assess the
reliability of the color catalogs being extracted from the EIS
multicolor data; 3) to provide ESO users with lists of potential VLT
targets.

In section 2, the basic characteristics of the color catalog relevant
to the present work are described.  In section 3, the color-color
diagram for point-like sources is inspected and empirical color
criteria are used to generate a preliminary list of candidate
white-dwarfs, low mass stars and quasars, with a range of
red-shifts. Results and the outlook for the future is presented in
section 4.

\section{The Point-Source Color Catalog}
\label{data}

The color catalog described here has been constructed from objects
detected in the 150 sec exposures using multicolor data presented in
paper~III. Color catalogs extracted from the co-added images will only
be available in the final release.

As discussed in paper~III, the generation of a color catalog for
generic use is a complex task given the range of possibilities in its
definition, which depends to a large extent on the science
goals. Since the intention of the present paper is primarily to
understand some general characteristics of the color catalog and
evaluate its usefulness for the primary science goals of the survey,
the following prescription has been adopted.  Starting from the unique
catalogs covering the whole patch extracted from the best available
images in each passband (see paper~III), a color catalog is created by
associating the objects detected in the different passbands (see
paper~I).  The final product is a unique entry for each object,
listing the photometric parameters and flags for each detection in
each passband, and information about the best seeing frames in all
bands. This information allows limits on the color for objects not
detected in some of the passbands to be calculated. At this point all
detections are considered, regardless of the band in which they were
detected and the nature of the object (star/galaxy).

It has been found, for instance, that this catalog contains objects
detected in the blue passbands without counterparts in $V$ and
$I$. Visual inspection of some of these cases, which represent roughly
2\% of the total number of objects for $B \lsim 22$, shows that the
unusual colors observed come mostly from objects detected at the
border of individual frames or along diffraction spikes of bright
stars. Others are due to ghost images present in the $B$ and $V$
images (see paper~III) and to merged images that are de-blended in
some passbands and not in others, depending on the seeing.  Even
though this is the most general color catalog, it is not the most
convenient catalog to use.  This is especially the case for extended
objects for which are contaminated by the problems mentioned above and
for which one would like to have the object centroid and measure
magnitudes with the same aperture in all bands. Another problem is
that the star/galaxy classification may vary from band to band, thus
implying that the object classification may not be unique.

To avoid some of these ambiguities, this paper considers only
point-sources as defined in the I-band. The selected sample includes
only objects detected in the $I$-band brighter than $I=23.0$ and with
SExtractor stellarity index $\geq 0.75$. In addition, all objects with
non-zero SExtractor and WeightWatcher flags (paper~I) are removed from
the catalog, in order to minimize contamination by spurious extreme
color objects, which originate from some of the cases described
above. It is worth pointing out that by eliminating all these objects
{a priori} may discard some interesting cases. The final sample is
available at ``http://www.eso.org/eis/''.

Note that star/galaxy classification is only reliable for magnitudes
brighter than $I \sim 21.5$ and colors were computed using the
mag\_auto estimator introduced in paper~I.  The mean error in the
colors are less than 0.15~mag for sources brighter than the
star-galaxy separation limit, and about 0.4~mag near the limiting
magnitude of the sample.

\begin{figure*}[ht]
\resizebox{0.75\hsize}{!}{\includegraphics{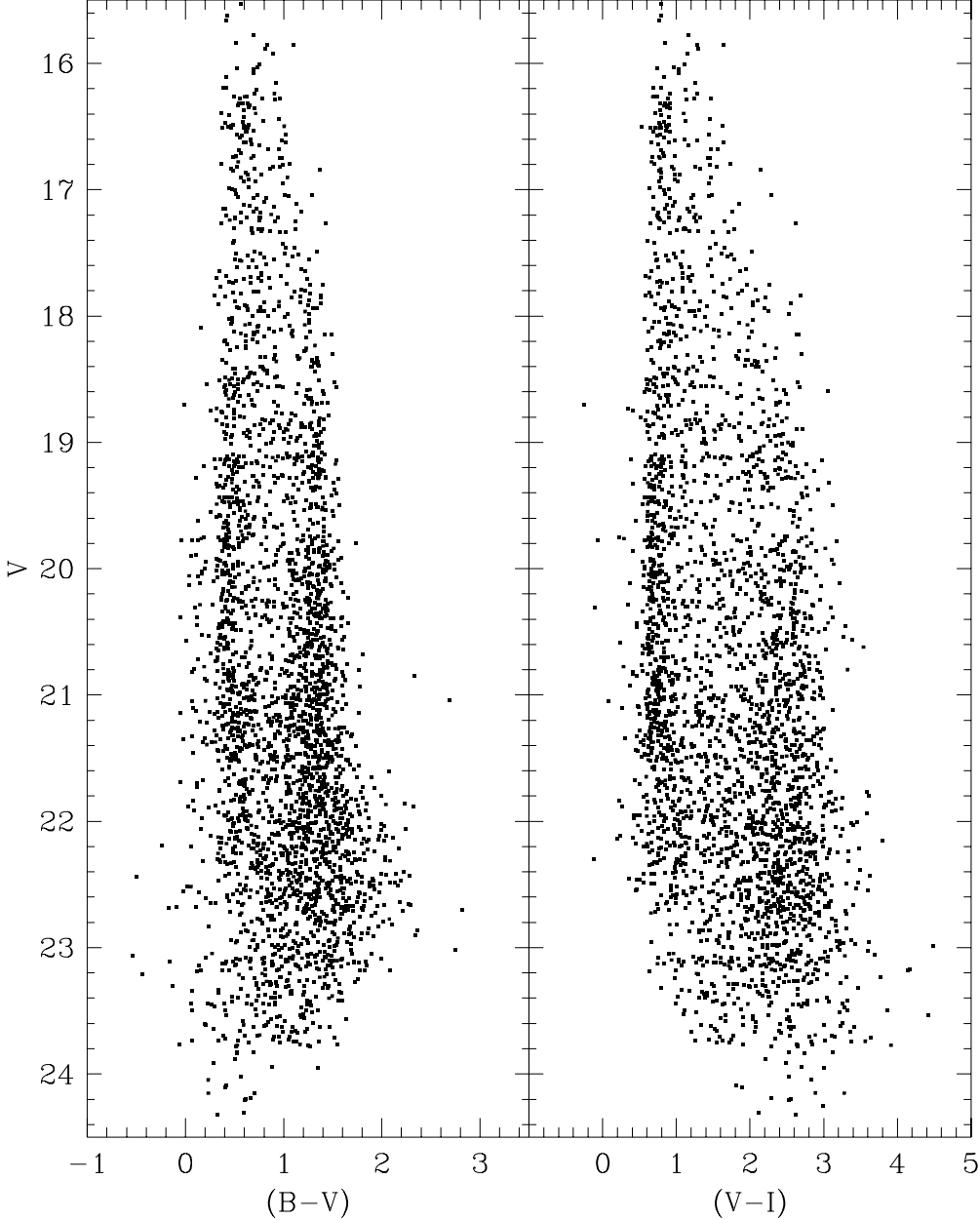}}               
\caption{Color-magnitude diagrams for EIS point-like sources.}
\label{fig:color-mag} \end{figure*}

Since the primary goal of this paper is to illustrate the possible use
of EIS data for different science goals and to evaluate the size of
interesting samples for follow-up work, the criteria adopted have been
in general conservative.  In particular, in the selection of targets
discussed below only reliable $I-$band detections, with
$\epsilon_I\lsim 0.2$ (S/N $\gsim$ 5) are considered.  Furthermore, to
avoid regions of very poor seeing and low transparency, some regions
were discarded, based on the distributions of seeing and limiting
isophothes shown in paper~III, leading to a sample covering a total
area of 1.27 square degrees (\eg figure~\ref{fig:starsdist}).

\begin{figure*}
\resizebox{0.49\hsize}{!}{\includegraphics{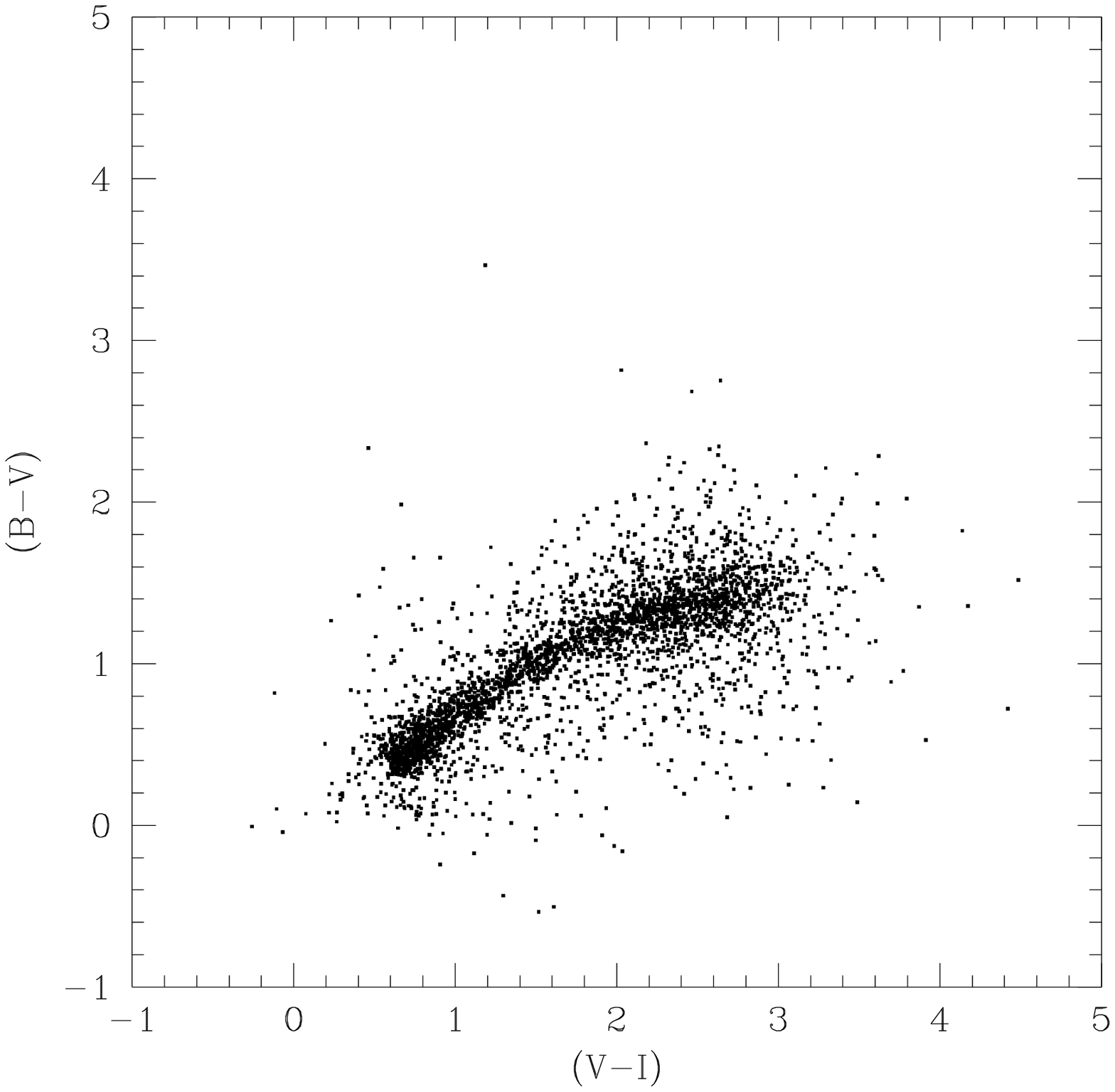}}               
\resizebox{0.49\hsize}{!}{\includegraphics{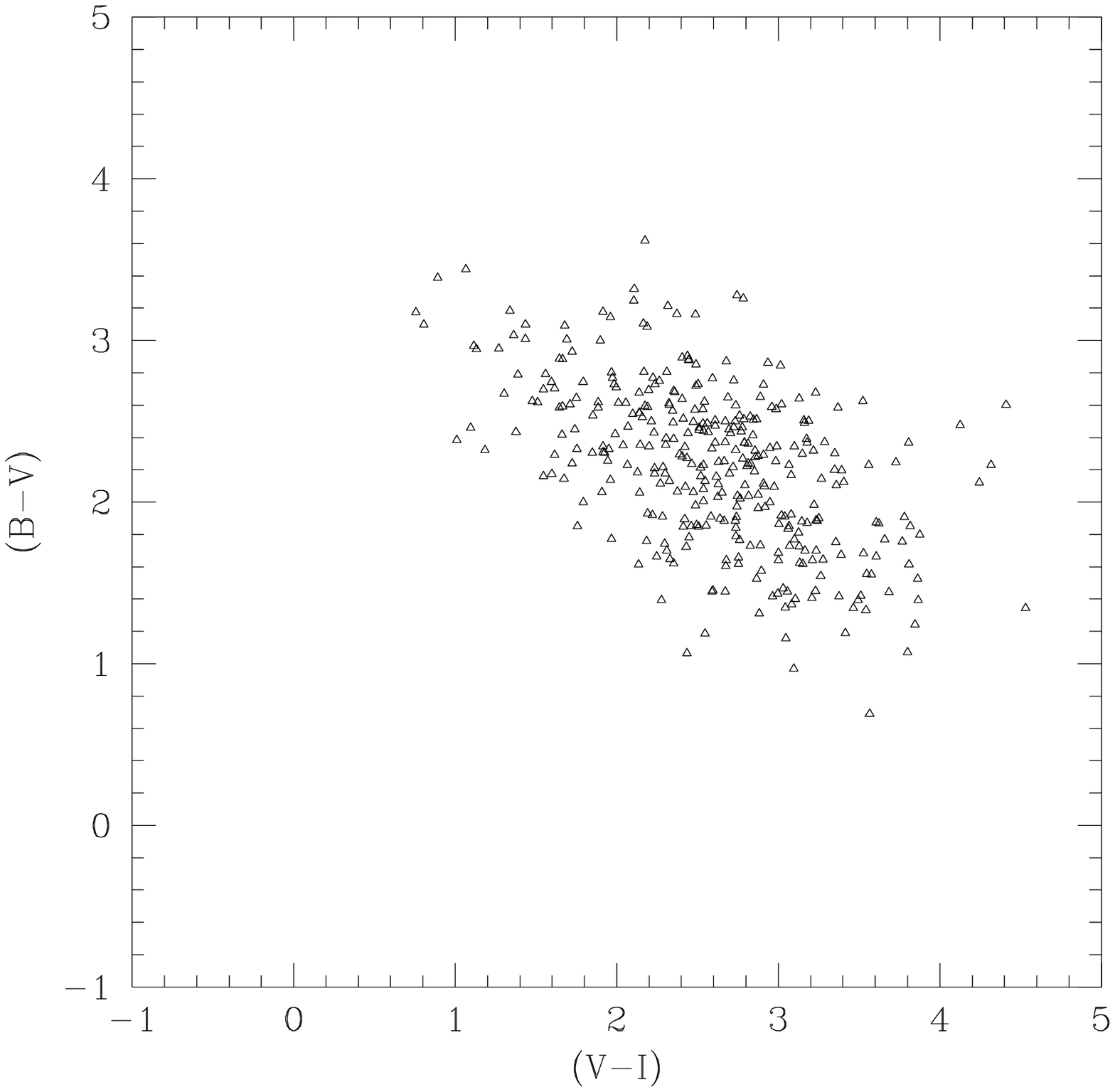}}               
\caption{Color-color diagram  for EIS point-sources. Left-panel:
those detected in all three passbands. Right-panel: those detected in
$V$ and $I$ but not $B$, for which the lower limit in $(B-V)$ is
indicated. }
\label{fig:colcolclean} 
\end{figure*}

To illustrate the general properties of the sample the color-magnitude
diagrams $V$ versus $(B-V)$ and $V$ versus $(V-I)$ are shown in
figure~\ref{fig:color-mag} for 3233 point-sources with S/N$\gsim$5 in
the $I$ passband. The color distribution for sources brighter
than $V=21$ has already been discussed in paper~III and compared with
model predictions. This plot is presented in V-band to make the
comparison with other works easier (\eg Reid and Majewski 1993). Note
that at faint magnitudes ($V~\gsim~21$) incompleteness in the stellar
sample sets in (see paper~III). In the figure blue, $(B-V)\sim 0.5$,
and red, $(B-V) \sim 1.5$ concentrations are clearly seen for $V\gsim
18.5$. This reflects the typical bimodal color distribution observed
for faint stars at high-galactic latitude, with the blue peak arising
from the turnoff of the main sequence for low-metalicity halo stars
and the red peak from disk stars. Note that the blue peak is
well-defined in the magnitude range $18.5~<~V~<~21.5$. For fainter
magnitudes it fades away partly due to the incompleteness of the
catalog and partly because at this magnitude limit one is approaching
the outer parts of the halo.

From the figure one sees that for $V\gsim19$ there is a large number
of objects bluer than the concentration associated with the halo
stars, and for $V\gsim21.5$ there is a population of very red objects
with $(V-I)>3.5$. Both are examples of interesting populations, the
identification of which is better explored using the color-color
diagram as discussed in the following section.

\section{Target Selection}
\label{target}

Figure~\ref{fig:colcolclean} shows the color-color diagrams for the
3233 objects detected (S/N $>$5) in the $I$ passband and the 345
objects detected in $V$ and $I$ only (S/N $>$5). The plot includes all
objects brighter than $I=23$. For non-detections in $B$ (hereafter
B-dropouts) an estimate of the $B$ limiting magnitude has been
measured on the best seeing frame. The limiting magnitude is defined
to be a 1$\sigma$ detection within the area corresponding to the
seeing-disk as measured in the $I$-band. From this an estimate of the
lower limit on the $(B-V)$ color is calculated. In addition to the
B-dropouts, there are objects only detected in the I-band, for which
the lower limit in $(V-I)$ is similarly computed. While a large number
of these objects is expected if one considers the sample as a whole
(because of the relative bright limiting magnitudes of the $V$
images), objects brighter than $I\sim 21$ are the most interesting and
are the ones considered in more detail below.

\begin{figure}[t]
\resizebox{\hsize}{!}{\includegraphics{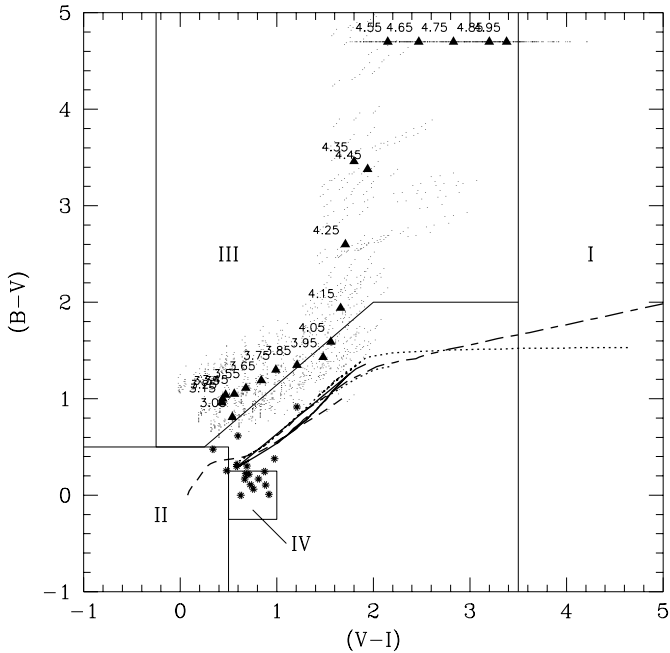}}               
\caption{Theoretical color-color plot for different type of objects.
The solid line shows the location of main sequence, sub- and red-giant
branch stars of an old halo, low metallicity, stellar population model
taken from Bertelli \etal (1994). The dotted line shows the location
of main sequence, sub- and red-giant branch stars of a young disk,
solar metallicity, stellar population model taken from Bertelli \etal
(1994).  The short-dashed line shows the location of a WD pure
Hydrogen cooling sequence taken from Bergeron, Wesemael, \& Beauchamp
(1995). The long-dashed line shows the location of 5~Gyr old BD stars
with solar metallicity, taken from Baraffe \etal (1998).  The color
track for QSOs at different redshifts ($3.05<z<5.00$) are shown by
triangles while the dots indicate the typical scatter around the
median for the different parameters of the spectral properties and
absorbers of high-redshift quasars (see text).  Also shown (stars) are
the EIS colors of the known quasars present in the EIS catalog which
have redshifts in the range $0.4<z<2.96$. }
\label{fig:coltheory} \end{figure}

For comparison with the previous figures, figure~\ref{fig:coltheory}
shows the locus of main sequence, giants, white dwarf and brown dwarf
stars. The stellar locus for main sequence, subgiant- and red-giant
branch stars typical of the old low-metallicity halo and the young
solar-type metallicity disk was taken from the models of Bertelli
\etal (1994) extending down to 0.6 $M_{\odot}$. The color-color
cooling sequence for pure-Hydrogen WD was taken from Bergeron,
Wesemael, \& Beauchamp (1995). Finally, the locus for very low mass
stars and/or brown dwarfs down to 0.08~$M_{\odot}$ is taken from the
models of Baraffe \etal (1998). This curves are presented in the
Jonhson-Cousins system, close to the EIS magnitude system except for
the $B-$band (paper~III). However, the differences are relatively
small and have no significant impact on the adopted selection criteria
described below.

Also shown in figure~\ref{fig:coltheory} is the track of quasars in
the color-color diagram as a function of redshift and the typical
color scatter along the sequence due to the different assumptions for
their typical spectra and intervening absorption.  QSO colors were
simulated using synthetic QSO spectra, which cover a range of
intrinsic spectral properties, and the response functions of the EIS
filters (paper~I). The method is the same as that used by Warren,
Hewett \& Osmer (1994) and Hall \etal (1996), and is a modified
version of the method of Warren \etal (1991). QSO spectra were
synthesised assuming that the QSO continuum has the form of a single
power law with spectral index $\alpha$
($\rm{S}(\nu)\propto\nu^{\alpha}$) and assuming fixed emission line
strengths relative to Ly$\alpha$+NV. Three different values of the
spectral index $\alpha=(-0.25, -0.75, -1.25)$ were used, and three
different values for the emission line strength, defined by the
Ly$\alpha$+NV rest-frame equivalent width, EW(Ly$\alpha$+NV)=(42, 84
and 168\AA). For each set of assumptions, spectra were generated at
intervals of 0.1 in $z$ over the range ($3.0 < z < 5.0$). Absorption
by intervening HI was taken into account by simulating absorption
spectra, following the method of Warren, Hewett \& Osmer (1994) and
based on the work of M{\o}ller and Jacobsen (1990).  For each set of
intrinsic properties, ten QSO spectra were generated at each $z$ step,
each using a different realization of the absorption spectrum
appropriate for that redshift.  Thus at each redshift a total of 90
spectra were generated.  Because patch~B is close to the South
Galactic Pole galactic extinction was neglected in the present
calculation.  Figure~\ref{fig:coltheory} shows the median and the
scatter corresponding to the various simulations as a function of
redshift.

\begin{table*}
\caption{Definition of regions of interest in figure~\ref{fig:coltheory} 
for candidate objects.}  
\label{tab:scheme}
\begin{tabular}{lll}
\hline \hline\noalign{\smallskip}
Region & Cand. Objects &Definition \\
\noalign{\smallskip}\hline\noalign{\smallskip}
I   & VLM/BD     & $(V-I)\geq 3.5$ \\
II  & WD         & $(V-I)\le0.5$ and $(B-V)\le$ 0.5 \\
\noalign{\medskip}
III & QSO        & $ \left\{ \begin{array}{l} 
-0.25<(V-I)<0.5\, {\rm and}\, (B-V)>0.5\\ 0.5<(V-I)<2.0\, {\rm and}\, (B-V)>0.86(V-I)+0.29  \\
2.0<(V-I)<3.5\, {\rm and}\, (B-V)>2.0  \end{array} \right. $ \\
\noalign{\medskip}
IV  & QSO Low z  & $0.5<(V-I)<1.0$ and $-0.25<(B-V)<0.25$ \\
\noalign{\smallskip}\hline \hline
  \end{tabular}
\end{table*}

In addition, in figure~\ref{fig:coltheory} all the 19 known quasars
present in the field are shown in their measured EIS magnitudes. These
quasars have redshifts, taken from the literature, in the range
$0.4<z<2.96$.

Comparison of the color-color diagram for the data and model
predictions shows at least four regions of potential interest. These
regions are schematically shown in figure~\ref{fig:coltheory} and
their limits are given in table~\ref{tab:scheme}. Objects in region~I
are candidate very low mass stars (VLM) or brown dwarf stars (BD),
those in region~II are candidate white dwarfs (WD).  Candidate quasars
(QSO) at different redshifts should lie in regions~III and IV.  Below
preliminary lists for these objects are presented in tables which
give: in column (1) the object name; in columns (2) and (3) the J2000
coordinates; in columns (4) and (5) the I magnitude and its error
estimate $\epsilon_{I}$; in columns (6) and (7) the (B-V) color and
its error estimate $\epsilon_{(B-V)}$; in columns (8) and (9) the
$(V-I)$ color and its error estimate $\epsilon_{(V-I)}$; and in column
(10) notes or comments on the individual objects, whenever necessary.
In the cases where the $(B-V)$ and/or $(V-I)$ colors are lower limits,
the measure is preceded by a $>$ sign and the error in the color is
the error in the magnitude in the passband in which the object is
detected. For objects not detected in two passbands the error in the
color is set to zero in the tables.

\subsection {Rare Stellar-type Candidates}

One of the interesting regions of the color-color diagram is the
region redder than $(V-I) \geq 3.5$ (region~I). Objects in this region
extend well beyond the track defined by main-sequence stars with
masses greater than $0.6M_{\odot}$.  Therefore, this region should be
populated primarily by very low mass stars ($0.6 > M/M_{\odot} > 0.1$)
in the disk and/or brown dwarfs.  Another possibility is that they are
asymptotic giant and red giant branch stars. However, this is unlikely
because there should be few of them in this color and magnitude range
since they would have to be high metallicity objects at very large
distances from the Sun ($\sim100$kpc). Even though unlikely,
considering the size of the area covered by the EIS multicolor data,
this region of the color space could also be populated by very
high-redshift QSOs with very large $(B-V)$, which could appear as $B$
non-detections. In this region there are 18 detections (listed in
table~\ref{tab:bdcand}); 22 B-dropouts with $(V-I)
\ge 3.5$, all brighter than $I=20$ (listed in table~\ref{tab:bdbdrop});
and 14 objects with $I\lsim21$, which are only detected in the I-band
(listed in table~\ref{tab:ionly}).  In the tables with ``rare''
stellar objects (\ref{tab:bdcand}, \ref{tab:bdbdrop}, and
\ref{tab:ionly}), the following naming convention has been adopted: 
VLM, for very low mass candidates, VLMB, for very low mass B-dropouts,
and VLMI, for the objects only detected in the $I-$band.

\begin{figure*}[ht]
\resizebox{0.9\hsize}{!}{\includegraphics{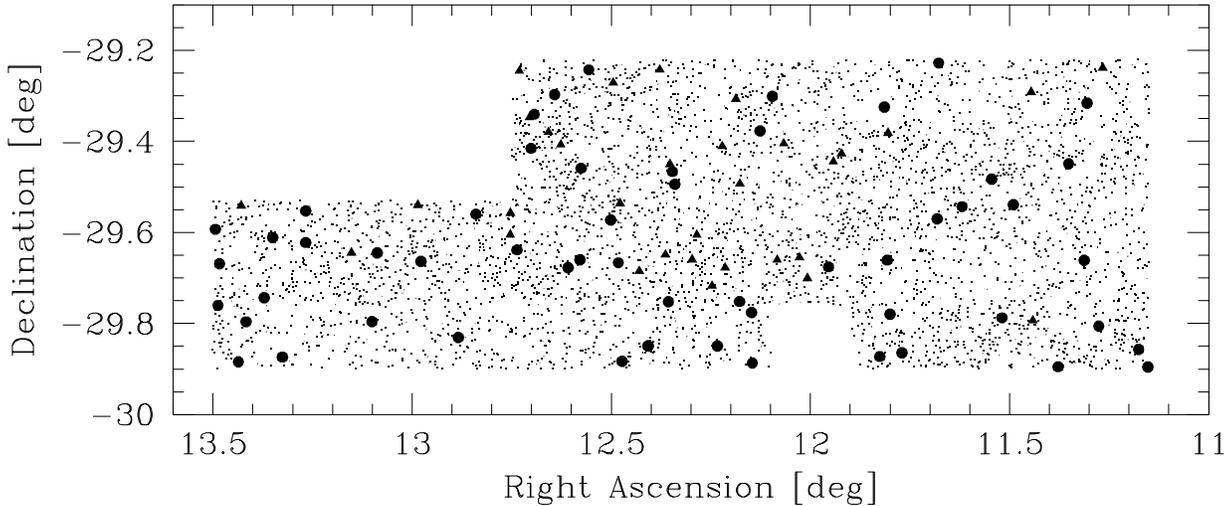}}
\caption{
Projected distribution of star-like objects which shows: all stellar
objects detected in the selected area of patch~B (dots); low-mass
candidates found in region~I of the color-color diagram of
figure~\ref{fig:coltheory} (filled circles); and WD candidates in
region~II (filled triangles)}
\label{fig:starsdist}
\end{figure*}

Since extreme colors could be caused by some unexpected artifact all
these cases have been visually inspected, and all seem to be
legitimate candidates. Note, however, that in the course of the
inspection the two brightest objects in this sample exhibited a
strange morphology in the coadded image appearing to be a ``double''
star, with the two objects having almost exactly the same magnitude,
$I=17.46 \pm0.01$, and a few arcsecs of separation. This prompted the
examination of the two single frames, which showed a single slightly
elongated object that occupies different positions in the two single
exposure images. The object was observed at $\alpha= 00^h 49^m
37^s.71$, $\delta= -29^\circ 50' 58''.7$, $JD=50696.3174202$ and
$\alpha= 00^h 49^m 37^s.76$, $\delta= -29^\circ 50' 56''.7$,
$JD=50696.32054438$. This fact strongly suggests that this object is
probably a relatively fast moving asteroid. However, no known
asteroids were found to be at the observed position during the nights
the observations were conducted.  This example of a serendipitous
source demonstrates the need to implement tools in the EIS pipeline to
search for transient phenomena present in the survey such as high
proper-motion objects, variables, supernovae.

%\subsection{White Dwarfs}

Another potentially interesting population is that defined by objects
in region~II of figure~\ref{fig:coltheory}. These objects are clearly
visible in figure~\ref{fig:color-mag} at magnitudes $V \gsim 19.5$.
These blue objects could be either relatively hot (young) disk white
dwarfs or blue horizontal branch (HB), low-metallicity halo
stars. However, for $V \gsim 20$ HB stars would be located at
$\gsim100$ kpc, where the density should be extremely small for
standard galactic structure models. There are 32 objects in region~II
which are listed in table~\ref{tab:wd}. The adopted cut-off in $(V-I)$
(see table~\ref{tab:scheme}) was chosen based on cooling sequence of
disk white dwarfs (Bergeron, Wesemael, \& Beauchamp 1995) shown in
figure~\ref{fig:coltheory}.  We emphasize that the criterion adopted
is somewhat arbitrary and it is used simply to illustrate the possible
identification of these objects. As can be seen from
figure~\ref{fig:coltheory}, this sample can be contaminated by low
redshift quasars. In fact table~\ref{tab:wd} contains 2 already known
quasar which are identified (name and redshift from the Simbad
database). The $U$-band data will be useful to sort out these cases.

Finally, figure~\ref{fig:starsdist} shows the spatial distribution of
these various candidates. Note that the northeast edge of the patch
has been removed because of the incompleteness of the B-band
catalogs. Similarly, a region along the southern edge was removed
because of the incompleteness in the I-band catalog. A small trimming
of the whole region has also been done yielding a total area of 1.27
square degrees.

\begin{figure*}
\resizebox{0.9\hsize}{!}{\includegraphics{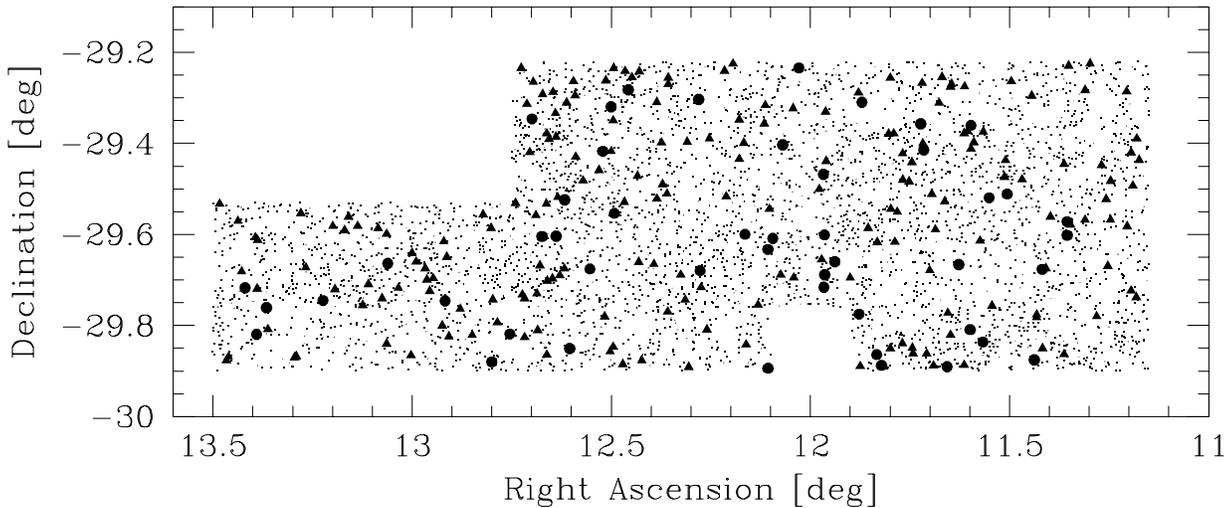}}
\caption{
Projected distribution of quasar candidates at low (filled circles),
intermediate and high redshift (filled triangles).  The adopted
selection criteria are discussed in the text.}
\label{fig:qsodist}
\end{figure*}

\subsection {Quasar Candidates}

 From simulations of QSO tracks (figure~\ref{fig:coltheory}) high
redshift QSOs ($3<z<5$) can be found in region~III of the color-color
diagram, while the available sample of known low redshift QSOs populate
region~IV (see figure~\ref{fig:coltheory}, Osmer \etal 1998). The
rough criteria used to define region~III (table~\ref{tab:scheme}) were
chosen based on the simulated QSO track. The blue part was chosen to
be parallel to the stellar locus but shifted to minimize the
contamination by stars. Several improvements in the selection can be
made to take into account the errors in colors, as a function of the
magnitude, and to optimize the yield based on the expected density of
objects of different types. Since the parent sample is public,
interested groups are likely to make significant refinements to the
selection criteria adopted here.

In region~III there are 70 objects detected in all three passbands.
These are listed in table~\ref{tab:qsocand}.  In addition, there are
126 objects that are detected in $V$ and $I$ but not detected in B
(hence have lower limits in $(B-V)$) that could also lie in
region~IV. These objects are listed in table~\ref{tab:qsobdrop}.  Note
that, since the depth of the B images varies across the patch, the
limits on $(B-V)$ are more meaningful in some areas than others.  The
depth of the $B$ frames corresponding to each object can be calculated
from the $V$ magnitude and the $(B-V)$ limits given in
table~\ref{tab:qsocand}.  In the tables the following naming
convention has been adopted: QSO and QSOB stand for objects in
region~III detected in all three bands ($\gsim3.0$) and B-dropouts
candidates, respectively.

Adopting the criteria given in table~\ref{tab:scheme} for region~IV,
where QSOs with $z\lsim3$ are likely to be found, one finds 48 stellar
objects which are listed in table~\ref{tab:qsolowz}. This table
includes 6 known QSOs, as indicated (name and redshift are from the
Simbad database). In the table QLZ stands for low redshift
($z\lsim3.0$) quasars. Note, however, that with the follow-up
observations in $U-$band to be carried out later this year, it will be
possible to select low-z QSOs more efficiently.

Figure~\ref{fig:qsodist} show the projected sky distribution of the
QSO candidates. This figure should be compared with those for the
seeing and the limiting magnitudes presented in paper~III to
investigate possible correlations between the QSO candidates and the
quality of the data, especially the B-dropouts or those detected only
in the $I$-band.  At first glance there is no obvious correlation as
the QSO candidates seem to be uniformly distributed over the surveyed
area.

\section{Summary and Outlook}

The ESO Imaging Survey is being carried out to help the selection of
targets for the first year of operation of VLT. This paper presents
some examples of possible candidates of interest, giving special
emphasis to stellar populations and quasars. Using the area covered by
the survey one is able to find candidate WDs and red objects likely to
be associated with very low mass stars or brown dwarfs. A preliminary
list is also presented for quasars. These lists and image postage
stamps in all three passbands are also available in the ESO Science
Archive server which allows the examination of the candidates
(``http://www.eso.org/eis/''). Finding charts can also be
easily extracted. Also available is the parent color sample from which
these candidates have been defined. It is important to emphasize that
in addition to providing these preliminary lists, the present work has
been an essential part in understanding the characteristics of the
color catalogs being produced and for the verification of their
reliability.

Improvements in the sample selection are certainly possible. Since the
data are publicly available, interested groups may refine the
selection criteria and produce their own samples.  The present results
lead to samples that are of the order of 50 to 100 candidates each.
The yield will only be defined by follow-up spectroscopic
observations. Much larger samples will be available from the Pilot
Survey to be carried out with the new wide-field camera (mecWIDE) on
the 2.2 m telescope at La Silla.

The present exploratory work has been done on a relatively small
sample with preliminary selection criteria and by handling the data in a
standard way. Nevertheless, it has already demonstrated the need for
providing users with more information than possible with traditional
catalogs, as emphasized by several other groups.  The exercise points
out the need for the development of an object-oriented database and
tools to inspect the multi-dimensional space of magnitudes, colors and
extraction parameters. In addition, full exploration of the data also
requires tools to facilitate the cross-identification of detected
objects with available databases in other wavelengths and with spectral
information, and to handle time-dependent information obtained in the
course of deep surveys.

The integration of EIS and the program being developed by the ESO
Science Archive Group is an essential step in the process of
translating results from multicolor deep (co-added) imaging surveys
into target lists for the VLT.  The Pilot Survey to be conducted on
the 2.2m telescope early next year offers the ideal test case to
define the basic science-driven requirements for data mining the
associated database. However, in order to take full advantage of the
new opportunities in a timely fashion, as required to support
VLT-science, a significant effort must be made well beyond what has
been done for EIS.  As generally recognized, the coming of age of
large, digital, multicolor imaging surveys creates new demands for
more efficient ways of extracting information. The development of
these tools and of efficient unsupervised pipelines is currently the
major challenge for the optimal extraction of valuable scientific
results.

\begin{acknowledgements}

We thank all the people directly or indirectly involved in the ESO
Imaging Survey effort. In particular, all the members of the EIS
Working Group for the innumerable suggestions and constructive
criticisms, the ESO Archive Group and the ST-ECF for their support.
Special thanks to A. Baker and D. Clements for their contribution in
the quasar search in the early stages of the EIS project. We would
also like to thank S. Warren for helpful comments and providing the
code for the calculation of the color track for quasars for the EIS
filters and I. Baraffe for providing the locus of brown dwarfs in the
appropriate passbands. Our special thanks to the efforts of
A. Renzini, VLT Programme Scientist, for his scientific input, support
and dedication in making this project a success. Finally, we would
like to thank ESO's Director General Riccardo Giacconi for making this
effort possible in the short time available.
This research has made use of the Simbad database, operated by the 
Centre de Données astronomiques de Strasbourg.
\end{acknowledgements}

\newpage
\newpage
\begin{table*}
\caption{Very Low-Mass Star Candidates.}  
\label{tab:bdcand}
\begin{tabular}{lcccccccc}
\hline \hline\noalign{\smallskip}
ID & \multicolumn{2}{c}{$\alpha$~~~~(J2000.0)~~~~$\delta$} & I & $\epsilon_{I}$ & (B-V)  & $\epsilon_{(B-V)}$ & (V-I)  & $\epsilon_{(V-I)}$\\
\noalign{\smallskip}\hline\noalign{\smallskip}
EIS VLM 1 & $00\, 45\, 30.77$ & $-29\, 53\, 41.8$ &  19.04 &   0.01 & $  1.82$ &   0.41 & $  4.14$ &   0.19  \\ 
EIS VLM 2 & $00\, 45\, 57.66$ & $-29\, 32\, 20.1$ &  19.36 &   0.02 & $  0.89$ &   0.41 & $  3.70$ &   0.26  \\ 
EIS VLM 3 & $00\, 47\, 13.69$ & $-29\, 39\, 38.1$ &  19.00 &   0.02 & $  1.36$ &   0.46 & $  4.17$ &   0.22  \\ 
EIS VLM 4 & $00\, 47\, 15.38$ & $-29\, 19\, 27.6$ &  18.17 &   0.01 & $  1.79$ &   0.29 & $  3.59$ &   0.08  \\ 
EIS VLM 5 & $00\, 47\, 49.07$ & $-29\, 40\, 32.3$ &  19.46 &   0.02 & $  0.96$ &   0.37 & $  3.77$ &   0.16  \\ 
EIS VLM 6 & $00\, 49\, 23.17$ & $-29\, 27\, 58.6$ &  19.11 &   0.02 & $  0.72$ &   0.44 & $  4.42$ &   0.19  \\ 
EIS VLM 7 & $00\, 49\, 53.47$ & $-29\, 52\, 58.9$ &  17.08 &   0.00 & $  1.55$ &   0.10 & $  3.54$ &   0.02  \\ 
EIS VLM 8 & $00\, 50\, 00.31$ & $-29\, 34\, 21.9$ &  19.33 &   0.01 & $  1.55$ &   0.42 & $  3.62$ &   0.15  \\ 
EIS VLM 9 & $00\, 50\, 18.26$ & $-29\, 27\, 31.2$ &  18.34 &   0.01 & $  1.59$ &   0.25 & $  3.59$ &   0.06  \\ 
EIS VLM 10 & $00\, 50\, 26.08$ & $-29\, 40\, 38.3$ &  20.07 &   0.03 & $  1.52$ &   0.46 & $  3.64$ &   0.25  \\ 
EIS VLM 11 & $00\, 50\, 56.76$ & $-29\, 38\, 17.5$ &  18.18 &   0.01 & $  1.99$ &   0.21 & $  3.62$ &   0.07  \\ 
EIS VLM 12 & $00\, 53\, 04.15$ & $-29\, 37\, 19.4$ &  18.50 &   0.01 & $  1.52$ &   0.29 & $  4.49$ &   0.13  \\ 
EIS VLM 13 & $00\, 53\, 18.17$ & $-29\, 52\, 25.2$ &  18.81 &   0.01 & $  2.28$ &   0.39 & $  3.62$ &   0.10  \\ 
EIS VLM 14 & $00\, 53\, 24.08$ & $-29\, 36\, 40.7$ &  19.86 &   0.02 & $  0.53$ &   0.34 & $  3.91$ &   0.22  \\ 
EIS VLM 15 & $00\, 53\, 44.66$ & $-29\, 53\, 03.5$ &  18.94 &   0.01 & $  1.58$ &   0.25 & $  3.60$ &   0.12  \\ 
EIS VLM 16 & $00\, 53\, 56.07$ & $-29\, 40\, 07.4$ &  19.42 &   0.02 & $  1.14$ &   0.35 & $  3.60$ &   0.26  \\ 
EIS VLM 17 & $00\, 53\, 57.00$ & $-29\, 45\, 36.7$ &  20.18 &   0.04 & $  1.13$ &   0.48 & $  3.56$ &   0.27  \\ 
EIS VLM 18 & $00\, 53\, 58.58$ & $-29\, 35\, 34.5$ &  19.62 &   0.02 & $  1.35$ &   0.49 & $  3.87$ &   0.29  \\ 
\hline \hline
\end{tabular}
\end{table*}

\begin{table*}
\caption{Very Low-Mass Star Candidates B-dropout.}      
\label{tab:bdbdrop}
\begin{tabular}{lcccccccc}
\hline \hline\noalign{\smallskip}
ID & \multicolumn{2}{c}{$\alpha$~~~~(J2000.0)~~~~$\delta$} & I & $\epsilon_{I}$ & (B-V)  & $\epsilon_{(B-V)}$ & (V-I)  & $\epsilon_{(V-I)}$ \\
\noalign{\smallskip}\hline\noalign{\smallskip}
EIS VLMB 1 & $00\, 44\, 42.23$ & $-29\, 51\, 24.4$ &  20.62 &   0.05 & $>  1.87$ &   0.43 & $  3.61$ &   0.44  \\ 
EIS VLMB 2 & $00\, 45\, 06.24$ & $-29\, 48\, 20.5$ &  20.73 &   0.05 & $>  1.76$ &   0.29 & $  3.77$ &   0.29  \\ 
EIS VLMB 3 & $00\, 45\, 13.27$ & $-29\, 18\, 57.3$ &  20.92 &   0.12 & $>  1.07$ &   0.47 & $  3.80$ &   0.48  \\ 
EIS VLMB 4 & $00\, 46\, 04.51$ & $-29\, 47\, 15.3$ &  20.19 &   0.03 & $>  2.25$ &   0.28 & $  3.73$ &   0.28  \\ 
EIS VLMB 5 & $00\, 46\, 10.71$ & $-29\, 28\, 58.7$ &  19.84 &   0.02 & $>  2.23$ &   0.17 & $  3.56$ &   0.17  \\ 
EIS VLMB 6 & $00\, 46\, 28.66$ & $-29\, 32\, 36.0$ &  20.16 &   0.03 & $>  1.77$ &   0.36 & $  3.66$ &   0.36  \\ 
EIS VLMB 7 & $00\, 46\, 43.65$ & $-29\, 34\, 13.2$ &  19.87 &   0.03 & $>  1.34$ &   0.37 & $  4.53$ &   0.37  \\ 
EIS VLMB 8 & $00\, 47\, 04.82$ & $-29\, 51\, 51.0$ &  19.43 &   0.02 & $>  2.23$ &   0.23 & $  4.32$ &   0.23  \\ 
EIS VLMB 9 & $00\, 48\, 30.31$ & $-29\, 22\, 37.1$ &  18.98 &   0.01 & $>  2.37$ &   0.11 & $  3.81$ &   0.11  \\ 
EIS VLMB 10 & $00\, 48\, 34.99$ & $-29\, 53\, 13.0$ &  20.47 &   0.05 & $>  1.42$ &   0.33 & $  3.51$ &   0.34  \\ 
EIS VLMB 11 & $00\, 48\, 35.42$ & $-29\, 46\, 32.0$ &  19.76 &   0.04 & $>  1.85$ &   0.25 & $  3.82$ &   0.26  \\ 
EIS VLMB 12 & $00\, 48\, 42.70$ & $-29\, 45\, 06.5$ &  20.76 &   0.08 & $>  1.44$ &   0.38 & $  3.69$ &   0.39  \\ 
EIS VLMB 13 & $00\, 49\, 21.62$ & $-29\, 29\, 37.8$ &  20.68 &   0.05 & $>  1.55$ &   0.27 & $  3.58$ &   0.27  \\ 
EIS VLMB 14 & $00\, 49\, 25.57$ & $-29\, 45\, 08.3$ &  20.88 &   0.06 & $>  1.66$ &   0.32 & $  3.61$ &   0.33  \\ 
EIS VLMB 15 & $00\, 50\, 18.77$ & $-29\, 39\, 34.8$ &  19.02 &   0.01 & $>  2.60$ &   0.19 & $  4.41$ &   0.19  \\ 
EIS VLMB 16 & $00\, 50\, 34.15$ & $-29\, 17\, 49.1$ &  18.49 &   0.01 & $>  2.48$ &   0.12 & $  4.12$ &   0.12  \\ 
EIS VLMB 17 & $00\, 50\, 46.36$ & $-29\, 20\, 25.1$ &  19.79 &   0.03 & $>  1.24$ &   0.21 & $  3.85$ &   0.21  \\ 
EIS VLMB 18 & $00\, 50\, 48.39$ & $-29\, 24\, 54.2$ &  19.40 &   0.02 & $>  2.62$ &   0.14 & $  3.52$ &   0.14  \\ 
EIS VLMB 19 & $00\, 51\, 21.59$ & $-29\, 33\, 35.3$ &  20.46 &   0.04 & $>  1.87$ &   0.25 & $  3.62$ &   0.25  \\ 
EIS VLMB 20 & $00\, 51\, 54.63$ & $-29\, 39\, 48.9$ &  20.40 &   0.04 & $>  1.80$ &   0.24 & $  3.88$ &   0.25  \\ 
EIS VLMB 21 & $00\, 53\, 04.07$ & $-29\, 33\, 08.7$ &  21.03 &   0.07 & $>  1.33$ &   0.43 & $  3.54$ &   0.43  \\ 
EIS VLMB 22 & $00\, 53\, 40.06$ & $-29\, 47\, 46.9$ &  20.30 &   0.04 & $>  1.91$ &   0.24 & $  3.78$ &   0.24  \\ 
\hline \hline
\end{tabular}
\end{table*}

\begin{table*}
\caption{Very Low-Mass Star Candidates, only detected in the $I-$band.}      
\label{tab:ionly}
\begin{tabular}{lcccccccc}
\hline \hline\noalign{\smallskip}
ID & \multicolumn{2}{c}{$\alpha$~~~~(J2000.0)~~~~$\delta$} & I & $\epsilon_{I}$ & (B-V)  & $\epsilon_{(B-V)}$ & (V-I)  & $\epsilon_{(V-I)}$\\
\noalign{\smallskip}\hline\noalign{\smallskip}
EIS VLMI 1 & $00\, 44\, 36.65$ & $-29\, 53\, 44.9$ &  20.91 &   0.06 & $>  0.56$ &   0.00 & $>  4.73$ &   0.06  \\ 
EIS VLMI 2 & $00\, 45\, 15.04$ & $-29\, 39\, 40.8$ &  20.77 &   0.08 & $>  0.39$ &   0.00 & $>  4.84$ &   0.08  \\ 
EIS VLMI 3 & $00\, 45\, 24.37$ & $-29\, 26\, 57.3$ &  20.76 &   0.05 & $>  0.32$ &   0.00 & $>  4.80$ &   0.05  \\ 
EIS VLMI 4 & $00\, 46\, 42.81$ & $-29\, 13\, 40.4$ &  20.87 &   0.08 & $>  0.43$ &   0.00 & $>  4.61$ &   0.08  \\ 
EIS VLMI 5 & $00\, 47\, 12.02$ & $-29\, 46\, 46.0$ &  20.55 &   0.05 & $>  0.39$ &   0.00 & $>  5.14$ &   0.05  \\ 
EIS VLMI 6 & $00\, 47\, 18.22$ & $-29\, 52\, 21.0$ &  20.81 &   0.06 & $>  0.44$ &   0.00 & $>  4.89$ &   0.06  \\ 
EIS VLMI 7 & $00\, 48\, 22.93$ & $-29\, 18\, 03.9$ &  20.72 &   0.06 & $>  0.06$ &   0.00 & $>  4.87$ &   0.06  \\ 
EIS VLMI 8 & $00\, 48\, 56.05$ & $-29\, 50\, 56.8$ &  20.58 &   0.14 & $> -0.26$ &   0.00 & $>  4.89$ &   0.14  \\ 
EIS VLMI 9 & $00\, 49\, 55.83$ & $-29\, 39\, 59.6$ &  20.89 &   0.06 & $>  0.37$ &   0.00 & $>  4.89$ &   0.06  \\ 
EIS VLMI 10 & $00\, 50\, 13.53$ & $-29\, 14\, 34.1$ &  20.25 &   0.04 & $> -0.23$ &   0.00 & $>  5.24$ &   0.04  \\ 
EIS VLMI 11 & $00\, 51\, 32.16$ & $-29\, 49\, 51.6$ &  20.98 &   0.06 & $>  0.54$ &   0.00 & $>  4.64$ &   0.06  \\ 
EIS VLMI 12 & $00\, 52\, 21.08$ & $-29\, 38\, 41.0$ &  20.90 &   0.06 & $>  0.37$ &   0.00 & $>  4.83$ &   0.06  \\ 
EIS VLMI 13 & $00\, 52\, 24.19$ & $-29\, 47\, 43.8$ &  20.86 &   0.06 & $>  0.53$ &   0.00 & $>  4.83$ &   0.06  \\ 
EIS VLMI 14 & $00\, 53\, 29.09$ & $-29\, 44\, 35.2$ &  20.02 &   0.03 & $>  0.55$ &   0.00 & $>  5.42$ &   0.03  \\ 
\hline\hline
\end{tabular}
\end{table*}

\newpage
\begin{table*}
\caption{Candidate White Dwarfs}
\label{tab:wd}
\begin{tabular}{lccccccccl}
\hline \hline\noalign{\smallskip}
ID & \multicolumn{2}{c}{$\alpha$~~~~(J2000.0)~~~~$\delta$} & I & $\epsilon_{I}$ & (B-V)  & $\epsilon_{(B-V)}$ & (V-I)  & $\epsilon_{(V-I)}$ & Notes\\
\noalign{\smallskip}\hline\noalign{\smallskip}
EIS WD 1 & $00\, 45\, 03.90$ & $-29\, 14\, 16.9$ &  18.39 &   0.02 & $  0.32$ &   0.01 & $  0.34$ &   0.02 & \\ 
EIS WD 2 & $00\, 45\, 45.94$ & $-29\, 47\, 36.4$ &  18.93 &   0.01 & $  0.32$ &   0.01 & $  0.40$ &   0.01 & \\ 
EIS WD 3 & $00\, 45\, 46.93$ & $-29\, 17\, 30.3$ &  20.55 &   0.07 & $  0.47$ &   0.05 & $  0.37$ &   0.07 & \\ 
EIS WD 4 & $00\, 47\, 13.29$ & $-29\, 22\, 52.4$ &  21.04 &   0.07 & $  0.20$ &   0.06 & $  0.30$ &   0.08 & \\ 
EIS WD 5 & $00\, 47\, 41.21$ & $-29\, 25\, 36.8$ &  21.55 &   0.09 & $  0.17$ &   0.10 & $  0.43$ &   0.10 & \\ 
EIS WD 6 & $00\, 47\, 46.22$ & $-29\, 26\, 39.5$ &  20.96 &   0.07 & $  0.39$ &   0.08 & $  0.48$ &   0.08 & \\ 
EIS WD 7 & $00\, 48\, 01.88$ & $-29\, 42\, 02.3$ &  19.47 &   0.02 & $  0.16$ &   0.02 & $  0.29$ &   0.02 & \\ 
EIS WD 8 & $00\, 48\, 06.62$ & $-29\, 39\, 18.0$ &  19.84 &   0.03 & $ -0.04$ &   0.01 & $ -0.07$ &   0.03 & \\ 
EIS WD 9 & $00\, 48\, 15.99$ & $-29\, 24\, 16.8$ &  19.46 &   0.02 & $  0.38$ &   0.02 & $  0.40$ &   0.02 & \\ 
EIS WD 10 & $00\, 48\, 20.09$ & $-29\, 39\, 36.7$ &  20.83 &   0.06 & $  0.08$ &   0.04 & $  0.27$ &   0.07 & \\ 
EIS WD 11 & $00\, 48\, 42.47$ & $-29\, 29\, 34.6$ &  18.34 &   0.01 & $  0.25$ &   0.01 & $  0.40$ &   0.01 & \\ 
EIS WD 12 & $00\, 48\, 44.90$ & $-29\, 18\, 24.0$ &  19.96 &   0.03 & $  0.47$ &   0.04 & $  0.47$ &   0.04 & \\ 
EIS WD 13 & $00\, 48\, 51.47$ & $-29\, 40\, 38.6$ &  21.62 &   0.13 & $  0.08$ &   0.08 & $  0.22$ &   0.14 & \\ 
EIS WD 14 & $00\, 48\, 53.04$ & $-29\, 24\, 37.5$ &  20.61 &   0.05 & $  0.40$ &   0.05 & $  0.48$ &   0.06 & \\ 
EIS WD 15 & $00\, 48\, 59.20$ & $-29\, 43\, 02.3$ &  17.89 &   0.00 & $  0.18$ &   0.01 & $  0.30$ &   0.01 & \\ 
EIS WD 16 & $00\, 49\, 08.36$ & $-29\, 36\, 18.4$ &  19.75 &   0.02 & $  0.45$ &   0.02 & $  0.37$ &   0.03 & \\ 
EIS WD 17 & $00\, 49\, 11.34$ & $-29\, 39\, 33.1$ &  19.17 &   0.01 & $  0.12$ &   0.01 & $  0.45$ &   0.02 & \\ 
EIS WD 18 & $00\, 49\, 24.64$ & $-29\, 26\, 59.4$ &  19.94 &   0.04 & $  0.28$ &   0.03 & $  0.34$ &   0.04 & \\ 
EIS WD 19 & $00\, 49\, 27.36$ & $-29\, 38\, 56.4$ &  21.88 &   0.13 & $  0.26$ &   0.11 & $  0.24$ &   0.15 & \\ 
EIS WD 20 & $00\, 49\, 30.88$ & $-29\, 14\, 33.6$ &  20.99 &   0.07 & $  0.33$ &   0.09 & $  0.44$ &   0.08 & \\ 
EIS WD 21 & $00\, 49\, 43.17$ & $-29\, 41\, 05.3$ &  20.49 &   0.05 & $  0.19$ &   0.04 & $  0.29$ &   0.05 & \\ 
EIS WD 22 & $00\, 49\, 54.72$ & $-29\, 32\, 10.0$ &  20.87 &   0.05 & $  0.07$ &   0.05 & $  0.46$ &   0.06 & \\ 
EIS WD 23 & $00\, 49\, 58.70$ & $-29\, 16\, 15.4$ &  19.53 &   0.02 & $  0.19$ &   0.02 & $  0.22$ &   0.02 & \\ 
EIS WD 24 & $00\, 50\, 30.48$ & $-29\, 24\, 24.8$ &  20.41 &   0.04 & $  0.10$ &   0.03 & $ -0.11$ &   0.04 & \\ 
EIS WD 25 & $00\, 50\, 37.90$ & $-29\, 22\, 45.6$ &  18.96 &   0.01 & $ -0.01$ &   0.01 & $ -0.26$ &   0.01 & ICS96 004812.3-293904\\ 
EIS WD 26 & $00\, 50\, 49.29$ & $-29\, 20\, 47.1$ &  20.65 &   0.06 & $  0.29$ &   0.10 & $  0.50$ &   0.07 & \\ 
EIS WD 27 & $00\, 50\, 55.45$ & $-29\, 14\, 40.6$ &  21.62 &   0.12 & $  0.02$ &   0.22 & $  0.27$ &   0.14 & \\ 
EIS WD 28 & $00\, 51\, 00.78$ & $-29\, 33\, 26.1$ &  18.75 &   0.01 & $  0.30$ &   0.01 & $  0.38$ &   0.01 & [CS83] 0048-2982, $z=2.439$ \\ 
EIS WD 29 & $00\, 51\, 00.94$ & $-29\, 36\, 15.6$ &  20.00 &   0.02 & $  0.38$ &   0.03 & $  0.46$ &   0.03 & \\ 
EIS WD 30 & $00\, 51\, 56.64$ & $-29\, 32\, 23.5$ &  20.97 &   0.06 & $  0.07$ &   0.04 & $  0.08$ &   0.06 & \\ 
EIS WD 31 & $00\, 52\, 36.64$ & $-29\, 38\, 40.6$ &  21.24 &   0.08 & $  0.18$ &   0.08 & $  0.44$ &   0.09 & \\ 
EIS WD 32 & $00\, 53\, 42.87$ & $-29\, 32\, 27.0$ &  20.42 &   0.04 & $  0.41$ &   0.04 & $  0.41$ &   0.04 & \\ 
\hline \hline
\end{tabular}
\end{table*}

\newpage
\begin{table*}
\caption{Quasar Candidates.}
%\label{tab:qsocand}
\begin{tabular}{lcccccccc}
\hline \hline\noalign{\smallskip}
ID & \multicolumn{2}{c}{$\alpha$~~~~(J2000.0)~~~~$\delta$} & I & $\epsilon_{I}$ & (B-V)  & $\epsilon_{(B-V)}$ & (V-I)  & $\epsilon_{(V-I)}$\\
\noalign{\smallskip}\hline\noalign{\smallskip}
EIS QSO 1 & $00\, 44\, 41.79$ & $-29\, 26\, 10.9$ &  21.78 &   0.13 & $  1.36$ &   0.29 & $  0.71$ &   0.16 \\ 
EIS QSO 2 & $00\, 44\, 43.33$ & $-29\, 23\, 22.3$ &  21.45 &   0.08 & $  1.48$ &   0.37 & $  1.14$ &   0.15 \\ 
EIS QSO 3 & $00\, 44\, 45.68$ & $-29\, 29\, 35.1$ &  21.60 &   0.10 & $  1.00$ &   0.21 & $  0.60$ &   0.13 \\ 
EIS QSO 4 & $00\, 44\, 46.71$ & $-29\, 43\, 25.5$ &  21.93 &   0.17 & $  1.59$ &   0.27 & $  0.56$ &   0.24 \\ 
EIS QSO 5 & $00\, 44\, 49.24$ & $-29\, 34\, 55.8$ &  20.15 &   0.04 & $  2.14$ &   0.29 & $  2.26$ &   0.11 \\ 
EIS QSO 6 & $00\, 44\, 59.01$ & $-29\, 28\, 56.7$ &  18.73 &   0.01 & $  2.17$ &   0.52 & $  3.49$ &   0.08 \\ 
EIS QSO 7 & $00\, 45\, 00.80$ & $-29\, 40\, 09.1$ &  20.68 &   0.07 & $  2.02$ &   0.30 & $  2.11$ &   0.15 \\ 
EIS QSO 8 & $00\, 45\, 01.81$ & $-29\, 31\, 20.5$ &  20.71 &   0.06 & $  1.92$ &   0.30 & $  1.80$ &   0.20 \\ 
EIS QSO 9 & $00\, 45\, 07.46$ & $-29\, 46\, 45.8$ &  21.85 &   0.12 & $  1.42$ &   0.38 & $  1.23$ &   0.25 \\ 
EIS QSO 10 & $00\, 45\, 14.51$ & $-29\, 34\, 05.4$ &  20.14 &   0.04 & $  2.08$ &   0.29 & $  2.35$ &   0.18 \\ 
EIS QSO 11 & $00\, 45\, 23.16$ & $-29\, 34\, 45.3$ &  20.70 &   0.06 & $  1.34$ &   0.15 & $  0.98$ &   0.11 \\ 
EIS QSO 12 & $00\, 45\, 24.25$ & $-29\, 13\, 45.4$ &  21.92 &   0.14 & $  1.24$ &   0.30 & $  0.74$ &   0.24 \\ 
EIS QSO 13 & $00\, 45\, 27.27$ & $-29\, 26\, 42.6$ &  21.79 &   0.11 & $  1.66$ &   0.39 & $  0.91$ &   0.18 \\ 
EIS QSO 14 & $00\, 45\, 40.05$ & $-29\, 51\, 01.0$ &  20.58 &   0.06 & $  1.47$ &   0.13 & $  0.53$ &   0.11 \\ 
EIS QSO 15 & $00\, 45\, 43.67$ & $-29\, 46\, 53.1$ &  21.67 &   0.11 & $  1.09$ &   0.19 & $  0.67$ &   0.18 \\ 
EIS QSO 16 & $00\, 45\, 43.68$ & $-29\, 46\, 34.4$ &  20.67 &   0.05 & $  2.82$ &   0.34 & $  2.03$ &   0.15 \\ 
EIS QSO 17 & $00\, 46\, 17.85$ & $-29\, 36\, 50.2$ &  21.35 &   0.09 & $  1.83$ &   0.37 & $  1.76$ &   0.23 \\ 
EIS QSO 18 & $00\, 46\, 27.15$ & $-29\, 16\, 30.6$ &  19.46 &   0.03 & $  2.03$ &   0.33 & $  2.58$ &   0.10 \\ 
EIS QSO 19 & $00\, 46\, 42.42$ & $-29\, 18\, 39.0$ &  19.02 &   0.03 & $  2.07$ &   0.29 & $  2.58$ &   0.08 \\ 
EIS QSO 20 & $00\, 46\, 44.64$ & $-29\, 35\, 20.1$ &  19.31 &   0.02 & $  2.04$ &   0.23 & $  2.56$ &   0.12 \\ 
EIS QSO 21 & $00\, 46\, 45.69$ & $-29\, 53\, 17.7$ &  20.38 &   0.05 & $  2.75$ &   1.04 & $  2.64$ &   0.27 \\ 
EIS QSO 22 & $00\, 46\, 46.81$ & $-29\, 30\, 40.5$ &  20.24 &   0.04 & $  2.18$ &   0.43 & $  2.39$ &   0.15 \\ 
EIS QSO 23 & $00\, 46\, 49.87$ & $-29\, 51\, 44.0$ &  22.10 &   0.18 & $  0.84$ &   0.22 & $  0.35$ &   0.26 \\ 
EIS QSO 24 & $00\, 46\, 58.17$ & $-29\, 51\, 41.1$ &  21.83 &   0.12 & $  1.99$ &   0.45 & $  0.67$ &   0.23 \\ 
EIS QSO 25 & $00\, 47\, 00.65$ & $-29\, 29\, 01.3$ &  20.81 &   0.06 & $  1.02$ &   0.15 & $  0.62$ &   0.10 \\ 
EIS QSO 26 & $00\, 47\, 04.48$ & $-29\, 50\, 19.8$ &  21.67 &   0.11 & $  1.37$ &   0.38 & $  0.99$ &   0.29 \\ 
EIS QSO 27 & $00\, 47\, 09.86$ & $-29\, 22\, 40.5$ &  21.73 &   0.10 & $  1.42$ &   0.26 & $  0.40$ &   0.13 \\ 
EIS QSO 28 & $00\, 47\, 11.39$ & $-29\, 32\, 38.8$ &  20.86 &   0.06 & $  1.05$ &   0.10 & $  0.62$ &   0.08 \\ 
EIS QSO 29 & $00\, 47\, 53.17$ & $-29\, 39\, 18.8$ &  18.89 &   0.01 & $  2.04$ &   0.22 & $  3.22$ &   0.08 \\ 
EIS QSO 30 & $00\, 47\, 54.67$ & $-29\, 30\, 01.2$ &  21.91 &   0.11 & $  1.13$ &   0.34 & $  0.91$ &   0.17 \\ 
EIS QSO 31 & $00\, 48\, 24.72$ & $-29\, 32\, 38.1$ &  20.41 &   0.04 & $  0.82$ &   0.04 & $  0.40$ &   0.04 \\ 
EIS QSO 32 & $00\, 48\, 51.66$ & $-29\, 14\, 26.5$ &  19.81 &   0.03 & $  2.08$ &   0.31 & $  2.50$ &   0.10 \\ 
EIS QSO 33 & $00\, 49\, 05.87$ & $-29\, 42\, 59.1$ &  19.74 &   0.02 & $  2.20$ &   0.26 & $  2.73$ &   0.11 \\ 
EIS QSO 34 & $00\, 49\, 15.26$ & $-29\, 44\, 37.8$ &  21.84 &   0.13 & $  1.59$ &   0.38 & $  1.52$ &   0.25 \\ 
EIS QSO 35 & $00\, 49\, 17.97$ & $-29\, 41\, 18.0$ &  20.57 &   0.05 & $  1.50$ &   0.17 & $  1.39$ &   0.09 \\ 
EIS QSO 36 & $00\, 49\, 26.05$ & $-29\, 46\, 09.4$ &  20.10 &   0.04 & $  2.03$ &   0.50 & $  2.20$ &   0.13 \\ 
EIS QSO 37 & $00\, 49\, 32.48$ & $-29\, 31\, 18.1$ &  19.41 &   0.02 & $  2.21$ &   0.38 & $  3.29$ &   0.14 \\ 
EIS QSO 38 & $00\, 49\, 52.12$ & $-29\, 31\, 42.8$ &  21.56 &   0.10 & $  1.28$ &   0.26 & $  1.01$ &   0.16 \\ 
EIS QSO 39 & $00\, 49\, 58.93$ & $-29\, 50\, 48.0$ &  18.57 &   0.01 & $  2.68$ &   0.29 & $  2.47$ &   0.04 \\ 
EIS QSO 40 & $00\, 50\, 04.00$ & $-29\, 46\, 48.0$ &  19.96 &   0.03 & $  2.11$ &   0.50 & $  2.86$ &   0.15 \\ 
EIS QSO 41 & $00\, 50\, 17.07$ & $-29\, 28\, 53.4$ &  19.40 &   0.02 & $  2.03$ &   0.37 & $  2.88$ &   0.09 \\ 
EIS QSO 42 & $00\, 50\, 26.94$ & $-29\, 18\, 36.4$ &  21.95 &   0.17 & $  0.50$ &   0.27 & $  0.19$ &   0.19 \\ 
EIS QSO 43 & $00\, 50\, 28.30$ & $-29\, 40\, 29.9$ &  20.66 &   0.05 & $  1.88$ &   0.36 & $  1.82$ &   0.12 \\ 
EIS QSO 44 & $00\, 50\, 32.21$ & $-29\, 31\, 04.0$ &  21.96 &   0.11 & $  0.96$ &   0.21 & $  0.50$ &   0.19 \\ 
EIS QSO 45 & $00\, 50\, 35.42$ & $-29\, 41\, 47.7$ &  21.10 &   0.07 & $  1.62$ &   0.26 & $  1.34$ &   0.17 \\ 
EIS QSO 46 & $00\, 50\, 38.87$ & $-29\, 22\, 42.3$ &  21.94 &   0.13 & $  1.21$ &   0.31 & $  0.75$ &   0.17 \\ 
EIS QSO 47 & $00\, 50\, 39.04$ & $-29\, 31\, 56.3$ &  21.45 &   0.10 & $  1.31$ &   0.33 & $  1.10$ &   0.18 \\ 
EIS QSO 48 & $00\, 50\, 42.90$ & $-29\, 40\, 04.8$ &  20.84 &   0.06 & $  2.08$ &   0.41 & $  2.34$ &   0.19 \\ 
EIS QSO 49 & $00\, 50\, 44.44$ & $-29\, 48\, 37.4$ &  20.33 &   0.04 & $  2.28$ &   0.44 & $  2.33$ &   0.17 \\ 
EIS QSO 50 & $00\, 51\, 12.46$ & $-29\, 35\, 08.7$ &  21.58 &   0.10 & $  1.72$ &   0.29 & $  1.22$ &   0.15 \\ 
EIS QSO 51 & $00\, 51\, 37.90$ & $-29\, 49\, 26.9$ &  21.73 &   0.11 & $  0.75$ &   0.26 & $  0.45$ &   0.27 \\ 
EIS QSO 52 & $00\, 51\, 39.14$ & $-29\, 38\, 57.7$ &  19.39 &   0.02 & $  2.16$ &   0.37 & $  3.11$ &   0.11 \\ 
EIS QSO 53 & $00\, 51\, 41.77$ & $-29\, 48\, 00.4$ &  20.35 &   0.04 & $  1.27$ &   0.05 & $  0.23$ &   0.05 \\ 
EIS QSO 54 & $00\, 51\, 47.19$ & $-29\, 41\, 45.3$ &  20.32 &   0.04 & $  2.05$ &   0.30 & $  2.11$ &   0.14 \\ 
EIS QSO 55 & $00\, 51\, 49.34$ & $-29\, 43\, 26.1$ &  20.03 &   0.03 & $  2.29$ &   0.40 & $  2.63$ &   0.13 \\ 
EIS QSO 56 & $00\, 51\, 51.09$ & $-29\, 41\, 58.0$ &  21.58 &   0.10 & $  1.72$ &   0.41 & $  1.54$ &   0.19 \\ 
EIS QSO 57 & $00\, 51\, 57.52$ & $-29\, 39\, 35.2$ &  20.27 &   0.04 & $  2.35$ &   0.55 & $  2.63$ &   0.19 \\ 
\hline \hline
\end{tabular}
\end{table*}

\newpage
\begin{table*}
\addtocounter{table}{-1}
\caption{Continued.}
\label{tab:qsocand}
\begin{tabular}{lcccccccc}
\hline \hline\noalign{\smallskip}
ID & \multicolumn{2}{c}{$\alpha$~~~~(J2000.0)~~~~$\delta$} & I & $\epsilon_{I}$ & (B-V)  & $\epsilon_{(B-V)}$ & (V-I)  & $\epsilon_{(V-I)}$\\
\noalign{\smallskip}\hline\noalign{\smallskip}
EIS QSO 58 & $00\, 52\, 00.10$ & $-29\, 38\, 26.4$ &  19.74 &   0.02 & $  2.22$ &   0.39 & $  2.66$ &   0.12 \\ 
EIS QSO 59 & $00\, 52\, 00.58$ & $-29\, 51\, 55.4$ &  19.94 &   0.03 & $  2.13$ &   0.38 & $  2.54$ &   0.23 \\ 
EIS QSO 60 & $00\, 52\, 15.43$ & $-29\, 50\, 24.9$ &  19.53 &   0.10 & $  2.12$ &   0.28 & $  2.61$ &   0.20 \\ 
EIS QSO 61 & $00\, 52\, 18.14$ & $-29\, 44\, 25.2$ &  19.42 &   0.02 & $  0.84$ &   0.03 & $  0.57$ &   0.02 \\ 
EIS QSO 62 & $00\, 52\, 20.53$ & $-29\, 35\, 11.1$ &  21.44 &   0.09 & $  1.17$ &   0.13 & $  0.51$ &   0.11 \\ 
EIS QSO 63 & $00\, 53\, 04.09$ & $-29\, 40\, 21.9$ &  19.64 &   0.02 & $  2.24$ &   0.35 & $  2.42$ &   0.08 \\ 
EIS QSO 64 & $00\, 53\, 07.12$ & $-29\, 33\, 13.0$ &  19.75 &   0.02 & $  2.12$ &   0.38 & $  2.73$ &   0.10 \\ 
EIS QSO 65 & $00\, 53\, 09.86$ & $-29\, 52\, 04.1$ &  19.26 &   0.02 & $  2.02$ &   0.27 & $  3.40$ &   0.12 \\ 
EIS QSO 66 & $00\, 53\, 42.79$ & $-29\, 40\, 53.4$ &  19.54 &   0.03 & $  2.23$ &   0.23 & $  2.32$ &   0.09 \\ 
EIS QSO 67 & $00\, 53\, 44.89$ & $-29\, 34\, 13.2$ &  20.68 &   0.05 & $  2.36$ &   0.50 & $  2.18$ &   0.20 \\ 
EIS QSO 68 & $00\, 53\, 50.63$ & $-29\, 52\, 06.5$ &  21.80 &   0.14 & $  1.56$ &   0.43 & $  1.48$ &   0.29 \\ 
EIS QSO 69 & $00\, 53\, 51.90$ & $-29\, 52\, 35.8$ &  19.47 &   0.02 & $  2.00$ &   0.26 & $  2.55$ &   0.11 \\ 
EIS QSO 70 & $00\, 53\, 55.95$ & $-29\, 31\, 57.9$ &  22.00 &   0.13 & $  1.05$ &   0.23 & $  0.46$ &   0.20 \\ 
\hline \hline
\end{tabular}
\end{table*}

\newpage
\begin{table*}
\caption{Quasar Candidates (B-dropouts).}      
%\label{tab:qsobdrop}
\begin{tabular}{lcccccccc}
\hline \hline\noalign{\smallskip}
ID & \multicolumn{2}{c}{$\alpha$~~~~(J2000.0)~~~~$\delta$} & I & $\epsilon_{I}$ & (B-V)  & $\epsilon_{(B-V)}$ & (V-I)  & $\epsilon_{(V-I)}$\\
\noalign{\smallskip}\hline\noalign{\smallskip}
EIS QSOB 1 & $00\, 44\, 43.60$ & $-29\, 44\, 15.6$ &  20.94 &   0.08 & $>  2.29$ &   0.41 & $  2.88$ &   0.41 \\ 
EIS QSOB 2 & $00\, 44\, 46.83$ & $-29\, 25\, 12.3$ &  20.84 &   0.05 & $>  2.43$ &   0.22 & $  2.70$ &   0.22 \\ 
EIS QSOB 3 & $00\, 44\, 49.53$ & $-29\, 17\, 09.5$ &  20.43 &   0.13 & $>  1.89$ &   0.29 & $  3.24$ &   0.31 \\ 
EIS QSOB 4 & $00\, 44\, 59.22$ & $-29\, 34\, 04.5$ &  20.10 &   0.04 & $>  2.51$ &   0.18 & $  3.16$ &   0.18 \\ 
EIS QSOB 5 & $00\, 45\, 04.54$ & $-29\, 26\, 52.9$ &  20.64 &   0.07 & $>  3.00$ &   0.20 & $  1.90$ &   0.21 \\ 
EIS QSOB 6 & $00\, 45\, 11.56$ & $-29\, 13\, 27.9$ &  19.87 &   0.03 & $>  3.26$ &   0.18 & $  2.79$ &   0.18 \\ 
EIS QSOB 7 & $00\, 45\, 14.48$ & $-29\, 16\, 58.2$ &  20.92 &   0.11 & $>  2.57$ &   0.30 & $  2.35$ &   0.32 \\ 
EIS QSOB 8 & $00\, 45\, 27.11$ & $-29\, 51\, 48.2$ &  20.81 &   0.05 & $>  2.11$ &   0.26 & $  3.36$ &   0.26 \\ 
EIS QSOB 9 & $00\, 45\, 35.39$ & $-29\, 33\, 39.5$ &  20.62 &   0.05 & $>  2.88$ &   0.22 & $  2.45$ &   0.23 \\ 
EIS QSOB 10 & $00\, 45\, 37.95$ & $-29\, 40\, 27.1$ &  20.94 &   0.12 & $>  2.46$ &   0.30 & $  2.51$ &   0.33 \\ 
EIS QSOB 11 & $00\, 45\, 46.59$ & $-29\, 17\, 44.2$ &  20.87 &   0.13 & $>  1.96$ &   0.27 & $  2.88$ &   0.30 \\ 
EIS QSOB 12 & $00\, 45\, 52.35$ & $-29\, 28\, 44.7$ &  20.33 &   0.03 & $>  2.58$ &   0.17 & $  2.99$ &   0.17 \\ 
EIS QSOB 13 & $00\, 45\, 59.03$ & $-29\, 15\, 50.4$ &  20.53 &   0.10 & $>  2.73$ &   0.19 & $  2.50$ &   0.22 \\ 
EIS QSOB 14 & $00\, 46\, 02.32$ & $-29\, 26\, 13.9$ &  20.58 &   0.04 & $>  1.89$ &   0.36 & $  3.24$ &   0.36 \\ 
EIS QSOB 15 & $00\, 46\, 03.03$ & $-29\, 28\, 24.6$ &  20.35 &   0.04 & $>  2.51$ &   0.16 & $  2.85$ &   0.16 \\ 
EIS QSOB 16 & $00\, 46\, 10.49$ & $-29\, 45\, 24.2$ &  20.82 &   0.08 & $>  1.98$ &   0.30 & $  3.22$ &   0.31 \\ 
EIS QSOB 17 & $00\, 46\, 15.88$ & $-29\, 22\, 26.7$ &  20.89 &   0.05 & $>  2.39$ &   0.15 & $  2.35$ &   0.16 \\ 
EIS QSOB 18 & $00\, 46\, 21.42$ & $-29\, 23\, 51.2$ &  20.90 &   0.05 & $>  1.69$ &   0.23 & $  3.00$ &   0.24 \\ 
EIS QSOB 19 & $00\, 46\, 23.17$ & $-29\, 24\, 42.5$ &  20.04 &   0.04 & $>  2.29$ &   0.14 & $  2.91$ &   0.14 \\ 
EIS QSOB 20 & $00\, 46\, 25.30$ & $-29\, 22\, 23.0$ &  20.77 &   0.05 & $>  2.68$ &   0.12 & $  2.14$ &   0.13 \\ 
EIS QSOB 21 & $00\, 46\, 27.31$ & $-29\, 22\, 40.1$ &  20.48 &   0.04 & $>  2.45$ &   0.14 & $  2.69$ &   0.14 \\ 
EIS QSOB 22 & $00\, 46\, 27.38$ & $-29\, 53\, 12.9$ &  20.41 &   0.04 & $>  2.64$ &   0.31 & $  3.13$ &   0.31 \\ 
EIS QSOB 23 & $00\, 46\, 35.01$ & $-29\, 16\, 10.1$ &  20.55 &   0.08 & $>  2.90$ &   0.17 & $  2.44$ &   0.18 \\ 
EIS QSOB 24 & $00\, 46\, 35.45$ & $-29\, 16\, 32.5$ &  20.15 &   0.07 & $>  2.87$ &   0.18 & $  2.68$ &   0.19 \\ 
EIS QSOB 25 & $00\, 46\, 35.55$ & $-29\, 49\, 14.4$ &  20.77 &   0.06 & $>  2.72$ &   0.24 & $  2.49$ &   0.25 \\ 
EIS QSOB 26 & $00\, 46\, 37.12$ & $-29\, 46\, 20.3$ &  20.81 &   0.06 & $>  3.25$ &   0.18 & $  2.11$ &   0.19 \\ 
EIS QSOB 27 & $00\, 46\, 39.14$ & $-29\, 31\, 37.9$ &  20.80 &   0.06 & $>  2.59$ &   0.24 & $  2.19$ &   0.25 \\ 
EIS QSOB 28 & $00\, 46\, 40.80$ & $-29\, 15\, 14.7$ &  20.24 &   0.08 & $>  2.75$ &   0.25 & $  2.73$ &   0.26 \\ 
EIS QSOB 29 & $00\, 46\, 48.08$ & $-29\, 40\, 43.9$ &  20.46 &   0.07 & $>  2.65$ &   0.20 & $  2.89$ &   0.21 \\ 
EIS QSOB 30 & $00\, 46\, 52.18$ & $-29\, 23\, 57.5$ &  20.94 &   0.06 & $>  2.55$ &   0.20 & $  2.10$ &   0.21 \\ 
EIS QSOB 31 & $00\, 46\, 52.77$ & $-29\, 16\, 02.0$ &  20.71 &   0.09 & $>  2.32$ &   0.24 & $  2.86$ &   0.25 \\ 
EIS QSOB 32 & $00\, 46\, 58.63$ & $-29\, 50\, 55.5$ &  20.80 &   0.08 & $>  2.10$ &   0.33 & $  2.97$ &   0.34 \\ 
EIS QSOB 33 & $00\, 46\, 58.86$ & $-29\, 26\, 28.7$ &  20.05 &   0.03 & $>  2.50$ &   0.16 & $  3.19$ &   0.16 \\ 
EIS QSOB 34 & $00\, 47\, 04.25$ & $-29\, 28\, 48.2$ &  20.89 &   0.05 & $>  2.24$ &   0.27 & $  2.81$ &   0.28 \\ 
EIS QSOB 35 & $00\, 47\, 04.44$ & $-29\, 25\, 18.8$ &  20.76 &   0.05 & $>  3.14$ &   0.16 & $  1.96$ &   0.16 \\ 
EIS QSOB 36 & $00\, 47\, 07.58$ & $-29\, 33\, 00.3$ &  20.62 &   0.05 & $>  1.87$ &   0.20 & $  3.18$ &   0.21 \\ 
EIS QSOB 37 & $00\, 47\, 09.19$ & $-29\, 36\, 58.3$ &  20.42 &   0.04 & $>  3.32$ &   0.19 & $  2.11$ &   0.19 \\ 
EIS QSOB 38 & $00\, 47\, 11.70$ & $-29\, 50\, 59.8$ &  20.46 &   0.04 & $>  2.49$ &   0.31 & $  3.16$ &   0.32 \\ 
EIS QSOB 39 & $00\, 47\, 11.71$ & $-29\, 15\, 21.8$ &  20.21 &   0.09 & $>  2.39$ &   0.27 & $  3.17$ &   0.28 \\ 
EIS QSOB 40 & $00\, 47\, 12.91$ & $-29\, 22\, 43.9$ &  20.91 &   0.06 & $>  2.52$ &   0.17 & $  2.41$ &   0.18 \\ 
EIS QSOB 41 & $00\, 47\, 19.76$ & $-29\, 37\, 01.4$ &  20.72 &   0.06 & $>  2.27$ &   0.26 & $  2.78$ &   0.26 \\ 
EIS QSOB 42 & $00\, 47\, 25.12$ & $-29\, 35\, 08.5$ &  20.79 &   0.06 & $>  1.88$ &   0.42 & $  3.15$ &   0.43 \\ 
EIS QSOB 43 & $00\, 47\, 29.97$ & $-29\, 53\, 22.3$ &  20.43 &   0.05 & $>  3.16$ &   0.23 & $  2.37$ &   0.23 \\ 
EIS QSOB 44 & $00\, 47\, 30.99$ & $-29\, 17\, 19.9$ &  20.31 &   0.08 & $>  2.77$ &   0.16 & $  2.59$ &   0.18 \\ 
EIS QSOB 45 & $00\, 47\, 35.96$ & $-29\, 41\, 42.4$ &  20.86 &   0.06 & $>  2.50$ &   0.19 & $  2.73$ &   0.20 \\ 
EIS QSOB 46 & $00\, 47\, 50.55$ & $-29\, 26\, 19.1$ &  20.70 &   0.07 & $>  2.11$ &   0.19 & $  2.91$ &   0.20 \\ 
EIS QSOB 47 & $00\, 47\, 50.86$ & $-29\, 19\, 49.7$ &  20.45 &   0.04 & $>  3.11$ &   0.11 & $  2.16$ &   0.11 \\ 
EIS QSOB 48 & $00\, 48\, 10.07$ & $-29\, 41\, 41.5$ &  20.21 &   0.04 & $>  3.62$ &   0.16 & $  2.17$ &   0.17 \\ 
EIS QSOB 49 & $00\, 48\, 10.48$ & $-29\, 19\, 21.9$ &  20.57 &   0.06 & $>  2.73$ &   0.12 & $  2.24$ &   0.14 \\ 
EIS QSOB 50 & $00\, 48\, 17.74$ & $-29\, 41\, 20.1$ &  20.77 &   0.06 & $>  2.51$ &   0.21 & $  2.79$ &   0.22 \\ 
EIS QSOB 51 & $00\, 48\, 27.26$ & $-29\, 18\, 57.3$ &  20.81 &   0.06 & $>  1.70$ &   0.24 & $  3.24$ &   0.25 \\ 
EIS QSOB 52 & $00\, 48\, 27.83$ & $-29\, 21\, 23.8$ &  20.57 &   0.05 & $>  2.51$ &   0.16 & $  2.61$ &   0.17 \\ 
EIS QSOB 53 & $00\, 48\, 31.62$ & $-29\, 45\, 15.9$ &  20.52 &   0.05 & $>  1.67$ &   0.32 & $  3.39$ &   0.32 \\ 
EIS QSOB 54 & $00\, 48\, 38.67$ & $-29\, 50\, 30.5$ &  20.30 &   0.05 & $>  2.10$ &   0.25 & $  2.91$ &   0.25 \\ 
EIS QSOB 55 & $00\, 48\, 40.04$ & $-29\, 23\, 56.0$ &  19.93 &   0.04 & $>  2.50$ &   0.10 & $  2.67$ &   0.10 \\ 
EIS QSOB 56 & $00\, 48\, 42.85$ & $-29\, 26\, 04.8$ &  20.03 &   0.04 & $>  2.60$ &   0.12 & $  2.74$ &   0.12 \\ 
EIS QSOB 57 & $00\, 48\, 42.93$ & $-29\, 20\, 52.4$ &  20.60 &   0.05 & $>  2.57$ &   0.22 & $  2.49$ &   0.23 \\ 
\hline \hline
\end{tabular}
\end{table*}

\newpage
\begin{table*}
\addtocounter{table}{-1}
\caption{Continued.}
%\label{tab:qsobdrop}
\begin{tabular}{lcccccccc}
\hline \hline\noalign{\smallskip}
ID & \multicolumn{2}{c}{$\alpha$~~~~(J2000.0)~~~~$\delta$} & I & $\epsilon_{I}$ & (B-V)  & $\epsilon_{(B-V)}$ & (V-I)  & $\epsilon_{(V-I)}$\\
\noalign{\smallskip}\hline\noalign{\smallskip}
EIS QSOB 58 & $00\, 48\, 46.70$ & $-29\, 13\, 31.0$ &  19.62 &   0.02 & $>  2.58$ &   0.14 & $  3.37$ &   0.14 \\ 
EIS QSOB 59 & $00\, 48\, 50.70$ & $-29\, 31\, 00.7$ &  20.96 &   0.06 & $>  2.60$ &   0.20 & $  2.33$ &   0.21 \\ 
EIS QSOB 60 & $00\, 49\, 00.87$ & $-29\, 23\, 20.1$ &  20.98 &   0.08 & $>  2.71$ &   0.14 & $  2.00$ &   0.16 \\ 
EIS QSOB 61 & $00\, 49\, 02.39$ & $-29\, 48\, 32.0$ &  20.91 &   0.11 & $>  3.18$ &   0.15 & $  1.92$ &   0.19 \\ 
EIS QSOB 62 & $00\, 49\, 13.25$ & $-29\, 53\, 29.1$ &  20.53 &   0.06 & $>  2.24$ &   0.27 & $  2.83$ &   0.28 \\ 
EIS QSOB 63 & $00\, 49\, 14.19$ & $-29\, 23\, 47.1$ &  19.64 &   0.03 & $>  2.46$ &   0.08 & $  2.73$ &   0.09 \\ 
EIS QSOB 64 & $00\, 49\, 25.56$ & $-29\, 16\, 11.0$ &  20.98 &   0.06 & $>  2.61$ &   0.19 & $  2.06$ &   0.20 \\ 
EIS QSOB 65 & $00\, 49\, 25.61$ & $-29\, 15\, 20.9$ &  20.19 &   0.04 & $>  2.23$ &   0.17 & $  3.07$ &   0.17 \\ 
EIS QSOB 66 & $00\, 49\, 26.46$ & $-29\, 30\, 36.8$ &  20.81 &   0.05 & $>  2.37$ &   0.26 & $  2.67$ &   0.26 \\ 
EIS QSOB 67 & $00\, 49\, 28.96$ & $-29\, 29\, 23.8$ &  20.27 &   0.03 & $>  2.37$ &   0.18 & $  3.18$ &   0.18 \\ 
EIS QSOB 68 & $00\, 49\, 29.81$ & $-29\, 23\, 53.5$ &  20.55 &   0.06 & $>  1.37$ &   0.20 & $  3.09$ &   0.21 \\ 
EIS QSOB 69 & $00\, 49\, 32.65$ & $-29\, 18\, 32.3$ &  20.25 &   0.04 & $>  2.41$ &   0.14 & $  2.84$ &   0.15 \\ 
EIS QSOB 70 & $00\, 49\, 34.33$ & $-29\, 39\, 53.7$ &  20.46 &   0.04 & $>  2.59$ &   0.19 & $  2.96$ &   0.20 \\ 
EIS QSOB 71 & $00\, 49\, 41.58$ & $-29\, 52\, 35.2$ &  20.42 &   0.04 & $>  2.88$ &   0.14 & $  2.44$ &   0.15 \\ 
EIS QSOB 72 & $00\, 49\, 43.18$ & $-29\, 14\, 30.5$ &  20.68 &   0.05 & $>  1.39$ &   0.42 & $  3.50$ &   0.43 \\ 
EIS QSOB 73 & $00\, 49\, 43.49$ & $-29\, 39\, 38.5$ &  20.58 &   0.05 & $>  2.14$ &   0.26 & $  3.27$ &   0.27 \\ 
EIS QSOB 74 & $00\, 49\, 44.25$ & $-29\, 28\, 16.9$ &  20.07 &   0.04 & $>  1.92$ &   0.19 & $  3.08$ &   0.19 \\ 
EIS QSOB 75 & $00\, 49\, 47.52$ & $-29\, 15\, 17.0$ &  20.36 &   0.04 & $>  2.81$ &   0.12 & $  2.31$ &   0.13 \\ 
EIS QSOB 76 & $00\, 49\, 50.13$ & $-29\, 16\, 56.0$ &  20.54 &   0.05 & $>  2.33$ &   0.17 & $  2.59$ &   0.17 \\ 
EIS QSOB 77 & $00\, 49\, 51.67$ & $-29\, 14\, 27.4$ &  20.57 &   0.05 & $>  1.85$ &   0.22 & $  3.07$ &   0.23 \\ 
EIS QSOB 78 & $00\, 49\, 52.98$ & $-29\, 53\, 06.2$ &  20.66 &   0.08 & $>  2.85$ &   0.25 & $  2.49$ &   0.26 \\ 
EIS QSOB 79 & $00\, 49\, 58.62$ & $-29\, 14\, 03.7$ &  20.99 &   0.07 & $>  1.85$ &   0.20 & $  2.51$ &   0.21 \\ 
EIS QSOB 80 & $00\, 49\, 58.73$ & $-29\, 20\, 59.2$ &  20.81 &   0.06 & $>  1.84$ &   0.20 & $  2.74$ &   0.21 \\ 
EIS QSOB 81 & $00\, 50\, 00.00$ & $-29\, 25\, 02.3$ &  20.14 &   0.04 & $>  1.70$ &   0.16 & $  3.17$ &   0.17 \\ 
EIS QSOB 82 & $00\, 50\, 00.68$ & $-29\, 51\, 25.5$ &  20.90 &   0.09 & $>  2.47$ &   0.31 & $  2.61$ &   0.32 \\ 
EIS QSOB 83 & $00\, 50\, 03.37$ & $-29\, 15\, 42.8$ &  20.24 &   0.04 & $>  2.49$ &   0.18 & $  2.53$ &   0.19 \\ 
EIS QSOB 84 & $00\, 50\, 07.37$ & $-29\, 27\, 32.0$ &  20.87 &   0.10 & $>  2.31$ &   0.12 & $  1.92$ &   0.15 \\ 
EIS QSOB 85 & $00\, 50\, 21.47$ & $-29\, 25\, 45.1$ &  20.46 &   0.04 & $>  2.37$ &   0.20 & $  2.79$ &   0.20 \\ 
EIS QSOB 86 & $00\, 50\, 21.86$ & $-29\, 17\, 41.1$ &  20.52 &   0.05 & $>  2.23$ &   0.16 & $  2.54$ &   0.17 \\ 
EIS QSOB 87 & $00\, 50\, 22.85$ & $-29\, 15\, 49.1$ &  19.56 &   0.02 & $>  2.32$ &   0.13 & $  3.22$ &   0.13 \\ 
EIS QSOB 88 & $00\, 50\, 31.00$ & $-29\, 41\, 24.1$ &  20.78 &   0.05 & $>  2.65$ &   0.21 & $  2.69$ &   0.22 \\ 
EIS QSOB 89 & $00\, 50\, 33.26$ & $-29\, 23\, 05.9$ &  19.61 &   0.02 & $>  2.68$ &   0.12 & $  3.23$ &   0.12 \\ 
EIS QSOB 90 & $00\, 50\, 33.58$ & $-29\, 20\, 01.3$ &  20.25 &   0.04 & $>  2.75$ &   0.11 & $  2.27$ &   0.12 \\ 
EIS QSOB 91 & $00\, 50\, 35.16$ & $-29\, 17\, 13.0$ &  20.90 &   0.07 & $>  2.07$ &   0.22 & $  2.38$ &   0.23 \\ 
EIS QSOB 92 & $00\, 50\, 37.43$ & $-29\, 23\, 27.0$ &  20.78 &   0.05 & $>  1.73$ &   0.24 & $  3.13$ &   0.25 \\ 
EIS QSOB 93 & $00\, 50\, 38.60$ & $-29\, 42\, 07.1$ &  20.50 &   0.04 & $>  2.61$ &   0.22 & $  3.02$ &   0.23 \\ 
EIS QSOB 94 & $00\, 50\, 38.82$ & $-29\, 51\, 50.7$ &  20.77 &   0.08 & $>  2.54$ &   0.29 & $  2.76$ &   0.30 \\ 
EIS QSOB 95 & $00\, 50\, 41.39$ & $-29\, 17\, 30.1$ &  20.14 &   0.05 & $>  2.30$ &   0.18 & $  2.39$ &   0.18 \\ 
EIS QSOB 96 & $00\, 50\, 44.83$ & $-29\, 43\, 47.3$ &  20.33 &   0.05 & $>  2.73$ &   0.29 & $  2.91$ &   0.29 \\ 
EIS QSOB 97 & $00\, 50\, 45.46$ & $-29\, 33\, 29.2$ &  20.31 &   0.03 & $>  2.37$ &   0.20 & $  3.29$ &   0.20 \\ 
EIS QSOB 98 & $00\, 50\, 46.94$ & $-29\, 15\, 53.2$ &  20.45 &   0.05 & $>  2.22$ &   0.16 & $  2.29$ &   0.17 \\ 
EIS QSOB 99 & $00\, 50\, 49.20$ & $-29\, 25\, 09.1$ &  20.03 &   0.03 & $>  3.28$ &   0.12 & $  2.75$ &   0.12 \\ 
EIS QSOB 100 & $00\, 50\, 50.95$ & $-29\, 18\, 47.6$ &  19.93 &   0.03 & $>  2.43$ &   0.10 & $  2.57$ &   0.11 \\ 
EIS QSOB 101 & $00\, 50\, 52.21$ & $-29\, 49\, 32.7$ &  20.81 &   0.07 & $>  1.90$ &   0.31 & $  3.25$ &   0.32 \\ 
EIS QSOB 102 & $00\, 50\, 52.22$ & $-29\, 44\, 25.6$ &  20.76 &   0.06 & $>  2.35$ &   0.22 & $  2.99$ &   0.23 \\ 
EIS QSOB 103 & $00\, 50\, 53.49$ & $-29\, 43\, 55.5$ &  21.00 &   0.07 & $>  2.62$ &   0.29 & $  2.54$ &   0.30 \\ 
EIS QSOB 104 & $00\, 50\, 54.29$ & $-29\, 14\, 03.6$ &  20.27 &   0.04 & $>  2.13$ &   0.19 & $  2.55$ &   0.20 \\ 
EIS QSOB 105 & $00\, 50\, 57.65$ & $-29\, 31\, 51.1$ &  20.93 &   0.05 & $>  2.37$ &   0.22 & $  2.79$ &   0.23 \\ 
EIS QSOB 106 & $00\, 51\, 08.70$ & $-29\, 47\, 35.6$ &  20.46 &   0.04 & $>  2.12$ &   0.37 & $  3.41$ &   0.37 \\ 
EIS QSOB 107 & $00\, 51\, 11.28$ & $-29\, 44\, 35.2$ &  20.77 &   0.07 & $>  2.36$ &   0.20 & $  2.81$ &   0.22 \\ 
EIS QSOB 108 & $00\, 51\, 17.08$ & $-29\, 33\, 22.8$ &  20.76 &   0.05 & $>  2.46$ &   0.21 & $  2.78$ &   0.21 \\ 
EIS QSOB 109 & $00\, 51\, 23.83$ & $-29\, 49\, 15.5$ &  20.77 &   0.05 & $>  2.51$ &   0.22 & $  2.87$ &   0.22 \\ 
EIS QSOB 110 & $00\, 51\, 31.02$ & $-29\, 45\, 49.3$ &  20.53 &   0.04 & $>  2.20$ &   0.36 & $  3.35$ &   0.37 \\ 
EIS QSOB 111 & $00\, 51\, 40.96$ & $-29\, 36\, 54.3$ &  20.67 &   0.05 & $>  3.09$ &   0.16 & $  2.19$ &   0.17 \\ 
EIS QSOB 112 & $00\, 51\, 52.19$ & $-29\, 40\, 27.9$ &  20.89 &   0.06 & $>  2.44$ &   0.23 & $  2.77$ &   0.23 \\ 
EIS QSOB 113 & $00\, 52\, 08.34$ & $-29\, 43\, 02.0$ &  20.78 &   0.06 & $>  2.25$ &   0.33 & $  2.98$ &   0.33 \\ 
EIS QSOB 114 & $00\, 52\, 15.47$ & $-29\, 35\, 56.2$ &  20.52 &   0.04 & $>  2.30$ &   0.32 & $  3.15$ &   0.32 \\ 
\hline \hline
\end{tabular}
\end{table*}

\newpage
\begin{table*}
\addtocounter{table}{-1}
\caption{Continued.}
\label{tab:qsobdrop}
\begin{tabular}{lcccccccc}
\hline \hline\noalign{\smallskip}
ID & \multicolumn{2}{c}{$\alpha$~~~~(J2000.0)~~~~$\delta$} & I & $\epsilon_{I}$ & (B-V)  & $\epsilon_{(B-V)}$ & (V-I)  & $\epsilon_{(V-I)}$ \\
\noalign{\smallskip}\hline\noalign{\smallskip}
EIS QSOB 115 & $00\, 52\, 26.34$ & $-29\, 42\, 32.6$ &  20.50 &   0.04 & $>  2.30$ &   0.37 & $  3.35$ &   0.37 \\ 
EIS QSOB 116 & $00\, 52\, 29.75$ & $-29\, 45\, 19.1$ &  20.38 &   0.05 & $>  3.22$ &   0.22 & $  2.32$ &   0.22 \\ 
EIS QSOB 117 & $00\, 52\, 32.78$ & $-29\, 34\, 53.2$ &  20.39 &   0.04 & $>  2.20$ &   0.24 & $  3.39$ &   0.25 \\ 
EIS QSOB 118 & $00\, 52\, 38.19$ & $-29\, 33\, 39.5$ &  20.80 &   0.05 & $>  2.28$ &   0.22 & $  2.86$ &   0.22 \\ 
EIS QSOB 119 & $00\, 52\, 40.58$ & $-29\, 35\, 26.8$ &  20.66 &   0.05 & $>  2.90$ &   0.24 & $  2.41$ &   0.25 \\ 
EIS QSOB 120 & $00\, 52\, 46.61$ & $-29\, 43\, 14.5$ &  20.47 &   0.04 & $>  3.16$ &   0.22 & $  2.49$ &   0.22 \\ 
EIS QSOB 121 & $00\, 52\, 47.74$ & $-29\, 34\, 53.8$ &  20.59 &   0.05 & $>  2.34$ &   0.19 & $  2.95$ &   0.20 \\ 
EIS QSOB 122 & $00\, 53\, 10.88$ & $-29\, 52\, 11.2$ &  20.26 &   0.04 & $>  2.86$ &   0.24 & $  2.94$ &   0.25 \\ 
EIS QSOB 123 & $00\, 53\, 26.89$ & $-29\, 48\, 29.2$ &  20.12 &   0.04 & $>  2.85$ &   0.20 & $  3.02$ &   0.20 \\ 
EIS QSOB 124 & $00\, 53\, 33.13$ & $-29\, 36\, 44.2$ &  20.87 &   0.06 & $>  2.69$ &   0.20 & $  2.35$ &   0.21 \\ 
EIS QSOB 125 & $00\, 53\, 33.35$ & $-29\, 43\, 08.5$ &  20.76 &   0.06 & $>  2.53$ &   0.21 & $  2.83$ &   0.22 \\ 
EIS QSOB 126 & $00\, 53\, 34.50$ & $-29\, 36\, 23.8$ &  20.70 &   0.05 & $>  2.17$ &   0.22 & $  3.08$ &   0.23 \\ 
\hline \hline
\end{tabular}
\end{table*}

\begin{table*}
\caption{Quasar Candidates (low redshift).}      
\label{tab:qsolowz}
\begin{tabular}{lccccccccl}
\hline \hline\noalign{\smallskip}
ID  & \multicolumn{2}{c}{$\alpha$~~~~(J2000.0)~~~~$\delta$}  & I & $\epsilon_{I}$ & (B-V)  & $\epsilon_{(B-V)}$ & (V-I)  & $\epsilon_{(V-I)}$ & Notes\\
\noalign{\smallskip}\hline\noalign{\smallskip}
EIS QLZ 1 & $00\, 45\, 25.31$ & $-29\, 36\, 07.1$ &  19.58 &   0.02 & $  0.12$ &   0.03 & $  0.87$ &   0.03 & \\ 
EIS QLZ 2 & $00\, 45\, 25.32$ & $-29\, 34\, 19.2$ &  20.70 &   0.07 & $ -0.02$ &   0.06 & $  0.65$ &   0.09 & \\ 
EIS QLZ 3 & $00\, 45\, 40.31$ & $-29\, 40\, 35.8$ &  19.68 &   0.03 & $  0.15$ &   0.02 & $  0.61$ &   0.04 & \\ 
EIS QLZ 4 & $00\, 45\, 45.07$ & $-29\, 52\, 32.3$ &  21.13 &   0.08 & $  0.23$ &   0.10 & $  0.76$ &   0.10 & \\ 
EIS QLZ 5 & $00\, 46\, 01.42$ & $-29\, 30\, 40.0$ &  20.15 &   0.04 & $  0.22$ &   0.04 & $  0.74$ &   0.05 & \\ 
EIS QLZ 6 & $00\, 46\, 12.25$ & $-29\, 31\, 10.3$ &  19.10 &   0.02 & $  0.11$ &   0.02 & $  0.79$ &   0.02 & ICS96 004345.8-294733\\ 
EIS QLZ 7 & $00\, 46\, 16.10$ & $-29\, 50\, 10.9$ &  17.86 &   0.00 & $  0.22$ &   0.01 & $  0.68$ &   0.01 & QSO 0043-3006, $z=1.124$\\ 
EIS QLZ 8 & $00\, 46\, 23.24$ & $-29\, 21\, 37.0$ &  20.35 &   0.04 & $  0.14$ &   0.05 & $  0.87$ &   0.05 & \\ 
EIS QLZ 9 & $00\, 46\, 23.78$ & $-29\, 48\, 32.9$ &  19.26 &   0.02 & $  0.17$ &   0.02 & $  0.78$ &   0.02 & \\ 
EIS QLZ 10 & $00\, 46\, 30.57$ & $-29\, 39\, 59.2$ &  20.51 &   0.07 & $  0.22$ &   0.06 & $  0.88$ &   0.08 & \\ 
EIS QLZ 11 & $00\, 46\, 37.56$ & $-29\, 53\, 25.2$ &  20.22 &   0.04 & $ -0.05$ &   0.05 & $  0.92$ &   0.06 & \\ 
EIS QLZ 12 & $00\, 46\, 51.71$ & $-29\, 24\, 52.5$ &  19.43 &   0.02 & $  0.15$ &   0.02 & $  0.68$ &   0.02 & \\ 
EIS QLZ 13 & $00\, 46\, 53.48$ & $-29\, 21\, 26.4$ &  21.64 &   0.10 & $  0.06$ &   0.17 & $  0.56$ &   0.12 & \\ 
EIS QLZ 14 & $00\, 47\, 17.09$ & $-29\, 53\, 15.4$ &  20.03 &   0.03 & $  0.09$ &   0.04 & $  0.76$ &   0.04 & \\ 
EIS QLZ 15 & $00\, 47\, 20.09$ & $-29\, 51\, 50.2$ &  22.54 &   0.19 & $  0.11$ &   0.28 & $  0.64$ &   0.26 & \\ 
EIS QLZ 16 & $00\, 47\, 28.91$ & $-29\, 18\, 34.7$ &  19.21 &   0.03 & $  0.12$ &   0.02 & $  0.57$ &   0.04 & \\ 
EIS QLZ 17 & $00\, 47\, 30.71$ & $-29\, 46\, 30.2$ &  17.42 &   0.00 & $  0.16$ &   0.00 & $  0.67$ &   0.01 & [CT83] 92, $z=2.021$\\ 
EIS QLZ 18 & $00\, 47\, 45.30$ & $-29\, 39\, 36.2$ &  19.43 &   0.02 & $  0.19$ &   0.02 & $  0.57$ &   0.02 & \\ 
EIS QLZ 19 & $00\, 47\, 51.33$ & $-29\, 41\, 18.2$ &  19.77 &   0.02 & $  0.23$ &   0.03 & $  0.76$ &   0.03 & \\ 
EIS QLZ 20 & $00\, 47\, 51.55$ & $-29\, 36\, 03.5$ &  18.53 &   0.01 & $  0.19$ &   0.01 & $  0.66$ &   0.01 & \\ 
EIS QLZ 21 & $00\, 47\, 51.92$ & $-29\, 42\, 58.8$ &  21.00 &   0.07 & $  0.12$ &   0.07 & $  0.73$ &   0.09 & \\ 
EIS QLZ 22 & $00\, 47\, 52.14$ & $-29\, 28\, 05.3$ &  20.50 &   0.05 & $  0.24$ &   0.06 & $  0.63$ &   0.05 & \\ 
EIS QLZ 23 & $00\, 48\, 06.89$ & $-29\, 14\, 03.2$ &  21.61 &   0.11 & $  0.20$ &   0.21 & $  1.00$ &   0.15 & \\ 
EIS QLZ 24 & $00\, 48\, 16.66$ & $-29\, 24\, 12.0$ &  20.96 &   0.07 & $  0.09$ &   0.11 & $  0.95$ &   0.08 & \\ 
EIS QLZ 25 & $00\, 48\, 22.62$ & $-29\, 36\, 30.8$ &  19.12 &   0.01 & $  0.06$ &   0.02 & $  0.77$ &   0.02 & \\ 
EIS QLZ 26 & $00\, 48\, 25.51$ & $-29\, 53\, 37.7$ &  19.37 &   0.02 & $  0.04$ &   0.03 & $  0.76$ &   0.03 & \\ 
EIS QLZ 27 & $00\, 48\, 25.67$ & $-29\, 37\, 58.5$ &  21.28 &   0.08 & $ -0.24$ &   0.11 & $  0.91$ &   0.12 & \\ 
EIS QLZ 28 & $00\, 48\, 39.50$ & $-29\, 35\, 58.9$ &  19.40 &   0.02 & $  0.12$ &   0.02 & $  0.66$ &   0.02 & \\ 
EIS QLZ 29 & $00\, 49\, 06.47$ & $-29\, 40\, 46.1$ &  20.00 &   0.03 & $  0.12$ &   0.03 & $  0.71$ &   0.04 & \\ 
EIS QLZ 30 & $00\, 49\, 07.34$ & $-29\, 18\, 12.6$ &  20.48 &   0.04 & $  0.22$ &   0.06 & $  0.74$ &   0.06 & \\ 
EIS QLZ 31 & $00\, 49\, 49.67$ & $-29\, 16\, 55.0$ &  19.71 &   0.02 & $  0.00$ &   0.04 & $  0.86$ &   0.03 & \\ 
EIS QLZ 32 & $00\, 49\, 58.15$ & $-29\, 33\, 12.6$ &  20.92 &   0.05 & $  0.12$ &   0.07 & $  0.77$ &   0.07 & \\ 
EIS QLZ 33 & $00\, 50\, 00.15$ & $-29\, 19\, 11.6$ &  21.17 &   0.08 & $  0.16$ &   0.15 & $  0.88$ &   0.11 & \\ 
EIS QLZ 34 & $00\, 50\, 05.11$ & $-29\, 25\, 02.1$ &  20.36 &   0.04 & $  0.06$ &   0.05 & $  0.77$ &   0.05 & \\ 
EIS QLZ 35 & $00\, 50\, 12.81$ & $-29\, 40\, 32.1$ &  20.15 &   0.03 & $  0.19$ &   0.04 & $  0.72$ &   0.04 & \\ 
EIS QLZ 36 & $00\, 50\, 25.10$ & $-29\, 51\, 01.7$ &  19.96 &   0.04 & $  0.07$ &   0.04 & $  0.86$ &   0.05 & \\ 
EIS QLZ 37 & $00\, 50\, 28.11$ & $-29\, 31\, 26.1$ &  19.41 &   0.02 & $  0.21$ &   0.02 & $  0.54$ &   0.02 & \\ 
EIS QLZ 38 & $00\, 50\, 33.18$ & $-29\, 36\, 13.7$ &  19.70 &   0.02 & $  0.11$ &   0.02 & $  0.68$ &   0.03 & \\ 
EIS QLZ 39 & $00\, 50\, 41.81$ & $-29\, 36\, 15.9$ &  18.56 &   0.01 & $  0.11$ &   0.01 & $  0.73$ &   0.01 & QSO 0048-298, $z=2.028$ \\ 
EIS QLZ 40 & $00\, 50\, 47.73$ & $-29\, 20\, 46.6$ &  20.44 &   0.07 & $  0.11$ &   0.10 & $  0.69$ &   0.08 & \\ 
EIS QLZ 41 & $00\, 51\, 01.05$ & $-29\, 49\, 07.2$ &  19.91 &   0.07 & $  0.17$ &   0.04 & $  0.62$ &   0.07 & \\ 
EIS QLZ 42 & $00\, 51\, 11.99$ & $-29\, 52\, 47.5$ &  19.52 &   0.02 & $  0.07$ &   0.03 & $  0.88$ &   0.03 & \\ 
EIS QLZ 43 & $00\, 51\, 40.09$ & $-29\, 44\, 45.7$ &  20.10 &   0.03 & $  0.20$ &   0.04 & $  0.94$ &   0.05 & \\ 
EIS QLZ 44 & $00\, 52\, 14.74$ & $-29\, 39\, 49.7$ &  20.85 &   0.06 & $  0.18$ &   0.06 & $  0.52$ &   0.08 & \\ 
EIS QLZ 45 & $00\, 52\, 53.76$ & $-29\, 44\, 42.2$ &  19.02 &   0.01 & $  0.06$ &   0.02 & $  0.76$ &   0.02 & QSO 0050-300, $z=1.922$\\ 
EIS QLZ 46 & $00\, 53\, 27.75$ & $-29\, 45\, 38.5$ &  19.21 &   0.02 & $  0.17$ &   0.02 & $  0.81$ &   0.02 & QSO 0051-300, $z=2.25$ \\ 
EIS QLZ 47 & $00\, 53\, 33.69$ & $-29\, 49\, 10.5$ &  20.67 &   0.05 & $  0.08$ &   0.06 & $  0.79$ &   0.07 & \\ 
EIS QLZ 48 & $00\, 53\, 40.66$ & $-29\, 43\, 01.3$ &  19.55 &   0.02 & $ -0.06$ &   0.03 & $  0.84$ &   0.03 & \\ 
\hline \hline
\end{tabular}
\end{table*}


\begin{thebibliography}{}

\bibitem[Baraffe98]{Baraffe98} 
Baraffe I., Chabrier G., Allard F., \& Hauschildt P. H., 1997, A\&A in
publication

\bibitem[Bertelli94]{1994A&AS..106..275B} 
Bertelli G., Bressan A., Chiosi C., Fagotto F., \& Nasi E., 1994,
A\&ASS, 106, 275

\bibitem[Bergeron95]{1995PASP..107.1047B} 
Bergeron P., Wesemael F., \& Beauchamp A., 1995, PASP, 107, 1047 

\bibitem{Boyle89} Boyle B.J. 1989, MNRAS, 240, 533

\bibitem[Hall96]{1996ApJ...462..614H} 
Hall P.B., Osmer P.S., Green R.F., Porter A.C., \& Warren S.J. 1996,
ApJ, 462, 614

\bibitem[Moller90]{Moller90} 
M{\o}ller P., Jakobsen P., 1990, A\&A, 228, 299

\bibitem[paperI]{paperI} 
Nonino M., Bertin E., da Costa L., Deul E., Erben T., Olsen L.,
Prandoni I., Scodeggio M., Wicenec A., Wichmann R., Benoist C.,
Freudling W., Guarnieri M. D., Hook I., Hook R., Mendez R., Savaglio
S., Silva S., Slijkhuis S., 1998, A\&A submitted, {\tt
astro-ph/9803336}, paper~I

\bibitem[Osmer98]{Osmer98} 
Osmer P.S., Kennefick J.D., Hall P.B., \& Green R.F., 1998, ApJ in
publication, {\tt astro-ph/9806366}

\bibitem[paperIII]{paperIII} 
Prandoni I., \etal, 1998, A\&A submitted, paper~III

\bibitem[RL97]{RL97} 
Renzini A., \& da Costa L., 1997, The Messenger, 87, 23

\bibitem[RM93]{RM93} 
Reid  N., \& Majewski, S.R. 1993, ApJ, 409, 635

\bibitem[Warren94]{Warren94} 
Warren S. J., Hewett P. C., Osmer P. S., 1994, ApJ, 421, 412

\bibitem[Warren91]{Warren91} 
Warren S. J., Hewett P. C., Irwin M. J., Osmer P. S., 1991, ApJS, 76, 1

\end{thebibliography}
\end{document}